\documentclass{aa} 
\usepackage{graphicx}
\usepackage{txfonts}
\usepackage{natbib,twoopt}
\bibpunct{(}{)}{;}{a}{}{,} %% natbib format for A&A and ApJ
\usepackage[breaklinks=true]{hyperref}

\begin{document}

%   \title{Asteroid 2023~DZ$_{2}$: A rapid characterization of a close passer (tentative title)}
   \title{Discovery and physical characterization as the first response to a potential asteroid
          collision: The case of 2023 DZ$_2$\thanks{Based on observations made with the Gran 
          Telescopio Canarias (GTC) telescope and the Isaac Newton Telescope (INT), in the 
          Spanish Observatorio del Roque de los Muchachos of the Instituto de Astrof\'{\i}sica de
          Canarias (program IDs GTC31-23A and INT99-MULTIPLE-2/23A), and the Two-Meter Twin 
          Telescope (TTT) and the Telescopio Carlos Sanchez (TCS), in the Spanish Observatorio del 
          Teide of the Instituto de Astrof\'{\i}sica de Canarias (commissioning phase).}}

%   \subtitle{I. Overviewing the $\kappa$-mechanism}

   \author{Marcel M. Popescu\inst{1}
          \and O. V\u{a}duvescu\inst{2,3,4}
          \and Julia de Le\'on \inst{3,5}
          \and C. de la Fuente Marcos\inst{6}
          \and R. de la Fuente Marcos \inst{7}
          \and M. O. St\u{a}nescu\inst{1}
          \and M. R. Alarcon\inst{3,5}
          \and M. Serra Ricart \inst{3, 5}
          \and J. Licandro\inst{3,5}
          \and D. Berte\c{s}teanu \inst{1}
          \and M. Predatu\inst{4}
          \and L. Curelaru \inst{1}
          \and F. Barwell \inst{2,8}
          \and K. Jhass \inst{2,8}
          \and C. Boldea \inst{4}
          \and A. Aznar Mac\'{i}as \inst{9}
          \and L. Hudin \inst{10} 
          \and B. A. Dumitru \inst{11}
          }
   \authorrunning{Popescu et al.}
   \titlerunning{Discovery and physical characterization of 2023 DZ$_2$}
   \offprints{Marcel M. Popescu, \email{popescu.marcel1983@gmail.com}}
   \institute{Astronomical Institute of the Romanian Academy, 5 Cuţitul de Argint, 040557 Bucharest, Romania
    \and  Isaac Newton Group of Telescopes (ING), Apto. 321, E-38700 Santa Cruz de la Palma, Canary    Islands, Spain
    \and Instituto de Astrof\'{\i}sica de Canarias (IAC), C/V\'{\i}a L\'{a}ctea s/n, 38205 La Laguna, Tenerife, Spain
    \and The University of Craiova, Str. A. I. Cuza nr. 13, 200585 Craiova, Romania
    \and Departamento de Astrof\'{\i}sica, Universidad de La Laguna, 38206 La Laguna, Tenerife, Spain
    \and Universidad Complutense de Madrid, Ciudad Universitaria, E-28040 Madrid, Spain
    \and AEGORA Research Group, Facultad de CC. Matem\'{a}ticas, Universidad Complutense de Madrid, Spain
    \and Department of Physics and Astronomy, University of Sheffield, Sheffield S3 7RH, UK
    \and Isaac Aznar Observatory, Alcublas, Valencia, Spain 
    \and ROASTERR-1 observatory, Cluj-Napoca, Romania
    \and Institute of Space Science (ISS), 409, Atomi\c{s}tilor Street, 077125 M\u{a}gurele, Ilfov, Romania
             }
   \date{Received 26 April 2023 / Accepted DD Mmmmm YYYY}

% \abstract{}{}{}{}{}
% 5 {} token are mandatory
 
  \abstract
   {Near-Earth asteroids (NEAs) that may evolve into impactors deserve 
    detailed threat assessment studies. 
    %Metallic asteroids may beat  current asteroid impact avoidance strategies and survive the fall through the atmosphere causing major damage when they hit our planet. 
    Early physical characterization of a would-be impactor may help in 
    optimizing impact mitigation plans. We first detected NEA 2023~DZ$_{2}$ 
    on 27--February--2023. After that, it was found to have a Minimum Orbit 
    Intersection Distance (MOID) with Earth of 0.00005~au as well as an 
    unusually high initial probability of becoming a near-term (in 2026) 
    impactor.}
   {We aim to perform a rapid but consistent dynamical and physical 
    characterization of 2023~DZ$_{2}$ as an example of a key response to mitigate the 
    consequences of a potential impact.}
   {We use a multi-pronged approach, drawing from various methods 
    (observational/computational) and techniques (spectroscopy/photometry 
    from multiple instruments), and bringing the data together to perform 
    a rapid and robust threat assessment.}
   {The visible reflectance spectrum of 2023~DZ$_{2}$ is consistent with 
    that of an X-type asteroid. Light curves of this object obtained on 
    two different nights give a rotation period $P$=6.2743$\pm$0.0005 min 
    with an amplitude $A$=0.57$\pm$0.14~mag. We confirm that although its 
    MOID is among the smallest known, 2023~DZ$_{2}$ will not 
    impact Earth in the foreseeable future as a result of secular
    near-resonant behaviour.}
   {Our investigation shows that coordinated observation and interpretation 
    of disparate data provides a robust approach from discovery to threat 
    assessment when a virtual impactor is identified. We prove that 
    critical information can be obtained within a few days after the announcement 
    of the potential impactor. }


   \maketitle
%
%________________________________________________________________

\section{Introduction}
When a new Solar system body (asteroid or comet) is discovered and its observations cover a short time interval, the orbit determination has a large uncertainty. Thus, at a given moment, the object can be anywhere inside a region defined by the propagation of the orbital uncertainties. \cite{2000Icar..145...12M} introduced the concept of virtual asteroid, a hypothetical object that follows any possible orbital solution obtained from the existing observations. If the dynamical evolution of any of these virtual asteroids is compatible with an Earth collision this is called a virtual impactor 
(abbreviated VI). Consequently, additional astrometric observations are needed to improve the orbit determination so the impact probability is better constrained, leading to its eventual removal as a threat (see e.g.  \citealt{2000Icar..145...12M, 2005dpps.conf..219M}).  This can be accomplished either by triggering new telescope observations or by data-mining the various observatories databases for "precoveries" (unnoticed apparitions in images acquired prior to its discovery) or both  (see e.g. \citealt{2013AN....334..718V,2020A&A...642A..35V}).

Large robotic surveys, such as the Panoramic Survey Telescope and Rapid Response System (Pan-STARRS, \citealt{2004SPIE.5489...11K, 2013PASP..125..357D}), the Catalina Sky Survey (CSS, \citealt{2012DPS....4421013C}), the Asteroid Terrestrial-impact Last Alert System (ATLAS, \citealt{2018PASP..130f4505T}), and the Zwicky Transient Facility (ZTF, \citealt{2019PASP..131g8002Y}) are continuously discovering most of the near-Earth asteroids (NEAs, or NEOs for near-Earth objects) down to an apparent $V$ magnitude as faint as 22 (depending on the survey). Nevertheless, there are regions of the sky which may not be observed for a long time by these facilities (due to the observing strategy, the weather, instrument updates, technical issues). Thus, the complementary observations performed by other facilities around the world are of significant importance for confirming, recovering, orbit refinement, and discovering NEOs.

The efficiency in finding unknown Solar system objects (SSOs) depends on two factors, the limiting magnitude that can be reached with an instrument and the sky area that can be covered. Because the majority of the facilities that aim at finding NEAs do not have access to telescopes with large aperture (compared with those of the above mentioned surveys), a wide field of view (FoV) covered during an observing run remains the accessible option. For example, an ingenious approach is used by Alain Maury, Georges Attard, Daniel Parrott, and their collaborators (the team of the MAP telescopes) to survey the Southern sky with small aperture telescopes having large FoV.\footnote{\url{https://www.spaceobs.com/en/Alain-Maury-s-Blog/2023-DW} accessed on 22-April-2023}

Close passers are small bodies that experience periodic close encounters with the Earth-Moon system and may eventually collide with Earth. Considering geological time scales, these small bodies represent a potential threat to life on our planet. This fact is evidenced by past impact events such as super-bolides, known craters, and mass extinctions attributed to major impacts. According to the most recent work, there are at least 210 known Earth craters caused by asteroids or comets \citep{2021M&PS...56.1024K,2022P&SS..22205575J}, while, on average, every year one meteoroid of about 3--4~m explodes in the atmosphere of our planet (see e.g. \citealt{2019MNRAS.483.5166D}).

The Chelyabinsk superbolide fall on 15-February-2013, reminded us that collisions of small Solar system  bodies with Earth may have catastrophic consequences \citep{2013Natur.503..238B, 2013Natur.503..235B}. Despite the relatively small size of this asteroid that entered through the atmosphere (about 18~m) on that date, it caused a significant amount of ground damage due to the air burst, affecting an area of more than 10\,000~$km^2$ and causing injuries to more than 1500 people (see e.g. \citealt{2018P&SS..160..107K}). This small NEA was not detected by active NEA surveys because its apparent motion was too close to the Sun prior to impact. Thus, no mitigation measurements were attempted.  

In order to mitigate these natural disasters, it is critical that these NEOs are first discovered with sufficient time prior to impact, and then their physical properties are determined as accurately and rapidly as possible. For example, metallic asteroids may overcome current asteroid impact avoidance strategies and survive after falling through the atmosphere causing major damage when they collide with our planet. The measures to prevent the impact or mitigate its consequences can only be taken based on information gained from detailed telescopic observations. The capability to offer a rapid response to a potential danger is a key factor for proper mitigation. Early physical characterization of a would-be impactor may help in  optimizing impact mitigation plans.

A recent example of this sequence of events at work is represented by 2023~DZ$_2$. We discovered this Apollo-class NEA on the night of 27-February-2023 using the 2.54~m Isaac Newton Telescope (INT)\footnote{\url{https://www.ing.iac.es/astronomy/telescopes/int/}} within the context of the EURONEAR (EUROpean Near Earth Asteroids Research) collaboration.\footnote{\url{http://www.euronear.org}} We promptly followed-up this discovery with astrometric observations during the next two nights. The data were submitted to the Minor Planet Center (MPC).\footnote{\url{https://www.minorplanetcenter.net/iau/mpc.html}} Immediately after the announcement of its discovery trough an Minor Planet Electronic Circular (MPEC 2023-F12, \citealt{2023MPEC....F...12B}), it was catalogued as a VI by Jet Propulsion Laboratory's (JPL) Sentry System for Earth impact monitoring,\footnote{\url{https://cneos.jpl.nasa.gov/sentry/}} by the NEODyS CLOMON2 Risk page\footnote{\url{https://newton.spacedys.com/neodys/index.php?pc=4.0}} list, and also by ESA Risk List.\footnote{\url{https://neo.ssa.esa.int/risk-list}} As additional observations were reported to the MPC by multiple observers around the world, the cumulative impact probability increased by several orders of magnitude (up to a cumulative impact probability of 0.0023 on March 18) but the analysis of the improved orbits led to the eventual removal of 2023~DZ$_2$ from the Sentry System on March 21. 
%__________________________________________________________________
%
%----------------------------------------------------------------------------------------------------------------------------------- TABLE I
%----------------------------------------------------------------------------------------------------------------- Orbital elements 2023 DZ2
%
   \begin{table}
    \centering
       \fontsize{8}{12pt}\selectfont
       \tabcolsep 0.14truecm
       \caption{\label{elements}Values of the heliocentric Keplerian orbital elements and their respective 1$\sigma$ uncertainties
                of 2023~DZ$_{2}$.
               }
       \begin{tabular}{lcc}
        \hline
         Orbital parameter                                 &   & value$\pm$1$\sigma$ uncertainty \\
        \hline
         Semimajor axis, $a$ (au)                          & = &   2.1555715$\pm$0.0000002        \\
         Eccentricity, $e$                                 & = &   0.53892721$\pm$0.00000005      \\
         Inclination, $i$ (\degr)                          & = &   0.0814345$\pm$0.0000012        \\
         Longitude of the ascending node, $\Omega$ (\degr) & = & 187.91380$\pm$0.00006            \\
         Argument of perihelion, $\omega$ (\degr)          & = &   5.95978$\pm$0.00006            \\
         Mean anomaly, $M$ (\degr)                         & = & 348.674236$\pm$0.000002          \\
         Perihelion distance, $q$ (au)                     & = &   0.993875393$\pm$0.000000007    \\
         Aphelion distance, $Q$ (au)                       & = &   3.3172677$\pm$0.0000003        \\
         Absolute magnitude, $H$ (mag)                     & = &  24.3$\pm$0.4                     \\
        \hline
       \end{tabular}
       \tablefoot{The orbit determination of 2023~DZ$_{2}$ is referred to epoch JD 2460000.5 
                  (25-Feb-2023) TDB (Barycentric Dynamical Time, J2000.0 ecliptic and equinox) and it is based on 635 observations with a data-arc span of 72 days (solution date, 24-April-2023, 08:41:00 PDT). The input data also include
                  radar observations (4 delay and 1 Doppler). Source: JPL's SBDB.
                 }
   \end{table}
%
%-------------------------------------------------------------------------------------------------------------------------------------------

Based on the apparent magnitude values reported together with the astrometric measurements by various observers, an absolute magnitude $H\approx$24 was estimated, which indicates a size in the range of 40~m to 100~m. A potential impact of such an asteroid can cause damages at local or regional level (see e.g. \citealt{1992spsu.conf.....M}). A close approach date of 25-March-2023 was estimated. As this NEA became brighter, more observations were reported including many by amateur astronomers who are able to acquire highly accurate data with affordable equipment (e.g. \citealt{2022PSJ.....3..156F,2022EPSC...16.1222N}). Fortunately, the impact probability decreased at insignificant levels ($\approx10^{-7}$) during the following days. The orbital elements of 2023~DZ$_{2}$ derived by using 634 observations with a data-arc covering of 72 days are reported in Table~\ref{elements}.  

Nevertheless, 2023~DZ$_2$ safely passed at a distance of 175\,030 km from Earth, on March 25 at 19:51 TDB 
(time-scale conversion difference TDB $-$ UT 69.185285~s), when it reached an apparent visual magnitude of 10.3 (for a minimum of $\approx$10~mag reached about two hours prior to perigee). Thanks to its brightness it offered a unique opportunity for characterization with various observational techniques: photometry, spectro-photometry, spectroscopy of various spectral intervals (visible, near-infrared, mid-infrared), polarimetry, and radar. 

The International Asteroid Warning Network (IAWN)\footnote{\url{https://iawn.net/}} organized a world wide campaign with the aim of involving as many observing facilities as possible, in a coordinated manner, to obtain the most accurate physical information about this object. This international organization was established in 2013 as a result of the UN-endorsed recommendations for an international response to a potential NEO impact threat. Their objective is to develop a "strategy using well-defined communication plans and protocols to assist Governments in the analysis of asteroid impact consequences and in the planning of mitigation responses." 

The close approach of 2023~DZ$_{2}$ offered a great opportunity for a world-wide collaboration to study a potential NEA impactor discovered one month prior to its close approach. Previous IAWN campaigns \citep{2019Icar..326..133R, 2022PSJ.....3..123R, 2022PSJ.....3..156F} targeted NEAs discovered some years before , thus allowing plenty of time to organize them. 

\begin{table}
\caption{Observational circumstances of 2023~DZ$_2$. .}
\label{tab:obs}
\centering
\begin{tabular}{cccccc}
\hline\hline\\[-3mm]
Obs. & Date & $m_V$ & $\alpha$ & $\Delta$ & $r$ \\
Type & Obs. (UTC) & & ($^{\circ}$) & (au) & (au) \\
\hline\\[-3mm]
Phot.  & 2023 03 20.9143 & 18.0 & 60.5 & 0.021 & 1.006\\
       & 2023 03 21.9284 & 17.5 & 60.9 & 0.017 & 1.004\\
       & 2023 03 22.9467 & 16.8 & 60.8 & 0.013 & 1.003\\
Colo.  & 2023 03 22.8725 & 17.0 & 60.7 & 0.013 & 1.003\\
Spec.  & 2023 03 17.8739 & 19.0 & 57.9 & 0.034 & 1.013\\
       & 2023 03 20.9161 & 18.0 & 60.5 & 0.021 & 1.006 \\
\hline
\end{tabular}
\tablefoot{Observation type includes time-series photometry (Phot.), colour photometry (Colo.) and visible spectra (Spec.). The UTC time corresponding to the start of the observations, the predicted apparent $V$ magnitude ($m_V$), the phase angle ($\alpha$), the geocentric($\Delta$) and heliocentric ($r$) distances are shown (these were obtained using the MPC ephemeris service accessed on 30-March-2023).}
\end{table}

In this paper, we show the results of our observing campaign, performed using several telescopes from the observatories of the Canary Islands and aimed at determining the physical properties of 2023~DZ$_2$. The objective of our work is to highlight the critical observational capabilities, both in terms of  instruments and data analysis resources, required to implement mitigation strategies to face the potential disasters coming from a cosmic hazard such as an asteroid impact. 

In less than a week since the announcement of the discovery and initial classification as VI of 2023~DZ$_2$, we were able to determine its spin rate and the light curve amplitude, its visible colours, and to obtain its visible spectrum. The observations performed are shown in Table~\ref{tab:obs}. Based on these observational data, we were able to constrain its size and estimate its composition. In addition, we were able to predict reliably its dynamical evolution in the time interval ($-$47, +142)~yr. This paper is organized as follows. In Section~2, we present the observations of 2023~DZ$_2$ and the methods used for its detection.  The dynamical evolution is discussed in Section~3. In Section~4, we present the photometric observations and the spin properties. The results of the analysis of the spectro-photometric data and the spectra are shown in Section~5. The discussion and the conclusions are shown in Section~6. Additional data obtained within the framework of our professional and amateur astronomers collaboration (the so-called ProAm) are discussed in Appendix~A.

\section{The detection of 2023~DZ$_2$}

The traditional way of finding new NEAs in CCD frames is the blinking detection technique. It begins with sequences of a few consecutive images of the same area of the sky, acquired with the same exposure parameters (exposure time, gain, etc.) during several tens of minutes. The raw images are processed and aligned using the stars in the field, so that any moving source (asteroid) appears to shift within the acquired image sequence. Assuming a linear motion in the short time interval covered by the observations, most of the noise can be identified and rejected, then the targets are validated, and their measurements reported to the MPC. In the early days of NEA surveys, a human operator manually performed most of these tasks. A popular representative example implementation of this technique, often applied two decades ago, involved the use of the Astrometrica software.\footnote{\url{http://www.astrometrica.at/}} With the advent of the all-sky surveys dedicated to the discovery of SSOs, more advanced detection software packages were developed \citep{2013PASP..125..357D}. 

Even the largest telescopes available cannot detect fainter and faster asteroids (as it is the case with most NEAs) through the blink method, due to atmospheric effects, mostly air glow (skyglow, light pollution, etc) and dispersion, and detector limitations. If a known object is invisible in the individual images, it is still possible to recover it through a technique called track-and-stack. In this case, multiple exposures are taken, which are shifted based on the predicted direction and speed of the target, then stacked to improve the signal to noise ratio, allowing the object to be detected with certainty in the co-added image. This technique makes sense for cases where the telescope does not support tracking an object itself (non-sidereal tracking) or when the trajectory is known only after acquiring the exposures. Otherwise, it is possible to just track according to the SSO apparent motion and then take a long exposure.

There is a natural extension of this method to the detection of unknown SSOs called Synthetic Tracking (also known as Digital Tracking). It was used to search for very faint Kuiper belt objects (KBOs) and trans-Neptunian objects (TNOs) which move very slowly \citep{1995ApJ...455..342C,1997A&A...317L..35G,1998AJ....116.2042G}. As suggested by its name, this method applies track-and-stack to "synthetic” motion vectors, typically all plausible motion vectors, while scanning for detections. However, practical synthetic tracking algorithms operate differently from track-and-stack --- the detection algorithm performs the stacking and it must identify automatically the unknown object, since the number of resulting stacked images is impossible to analyze in practical time by human validators.

The key factor for all these techniques is the time needed for the data reduction, because the small close-approaching NEAs are only observable (at apparent magnitudes accessible with ground-based telescopes) for very short time intervals. While modern instruments cover larger fields of view (already of the order of several square degrees) with fainter magnitude limits, the effort needed to reduce the amount of data collected becomes prohibitive for small research groups or amateurs. Thus, fast and accurate algorithms are required for complementing the ongoing robotic all-sky surveys carried out by large research consortia (e.g. Pan-STARRS, CSS, ATLAS, ZTF). One successful example of such a software application is Tycho Tracker\footnote{\url{https://www.tycho-tracker.com/}} that is used by both amateur and professional astronomers. This software provides an easy-to-use interface which facilitates the detection and measurement of asteroids, comets, and variable stars. It also supports the synthetic tracking technique. Nevertheless, the integration of this software in an automatic data reduction pipeline is not straightforward, and the computation power required to search for fast moving NEAs in a large field of view increases exponentially.

Within the EURONEAR collaboration, we started the development of STU (Synthetic Tracking on Umbrella), a synthetic tracking pipeline that aims to run in "near-real time" with the observations. It is a new addition to the Umbrella software suite \citep{2021A&C....3500453S} and it uses OpenCL to tap into the power of modern GPUs. As a pipeline, STU begins with image conditioning tasks, such as star masking (median stacking and star detection/mask generation through Umbrella2, handled on the CPU, masking the original images through OpenCL on the GPU). This is followed by re-projecting the input images into a common gnomonic projection chart (so that asteroids move, to first approximation, in straight lines in X-Y coordinates), directly on the GPU \citep{stuacm2013proceedings}.

The data cube is then processed in "scan stage", which is the actual GPU-based search. The search algorithm, applies a combination of threshold counting \citep{2021PASJ...73..519Y} that is a generalization of thresholding the median value that is very robust against false positives, with the more common addition described by \citet{2019AJ....157..119W} for refining the results (since computing the median is expensive, and threshold counting does not provide additional information for bright targets).

\begin{table*}
\caption{Observational circumstances for the detection of 2023~DZ$_2$. }
\label{tab:astrom}
\centering
\begin{tabular}{c c c c c c c}
\hline\hline\\[-3mm]
Field name & Date & UT start & UT end & Nexp & AM & Seeing \\
\hline\\[-3mm]
n1o1    & 27-Feb-2023 & 22:21 & 23:31 & 12 & 1.02 & 1.5 \\
E309252 & 28-Feb-2023 & 22:14 & 22:28 & 12 & 1.01 & 1 \\
E309252 & 01-Mar-2023 & 22:46 & 22:00 & 12 & 1.02 & 1.1 \\
\hline
\end{tabular}
\tablefoot{The date and Universal Time (UT) for the beginning and for the end of the observing set, the number of exposures (NExp), the mid-observation  airmass (AM), and the median seeing (Seeing) in arcseconds are shown. }
\end{table*}

Thus, we first detected 2023~DZ$_{2}$ \citep{2023MPEC....F...12B} thanks to the real-time processing using the software tools and infrastructure of the Romanian ParaSOL project.\footnote{\url{https://planet.astro.ro/ParaSOL/}} Table~\ref{tab:astrom} summarizes the observing log. The objective of the observing run was to prove the near-real time capabilities of the developed software, including both STU, the sensor correction and plate solving pipeline. Nevertheless, this serendipitous discovery was made within the framework of our observing campaigns whose aim is finding new asteroids and comets (NEAs, Atiras and Vatiras). The observing strategy was designed to cover the sky regions where it is more likely to detect unknown NEAs (by exploring ranges of ecliptic longitude neglected by the main surveys).

\begin{figure*}[h!]
    \begin{center}
    \includegraphics[width=6cm]{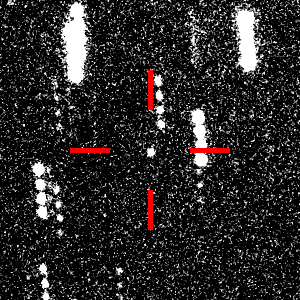}
    \includegraphics[width=6cm]{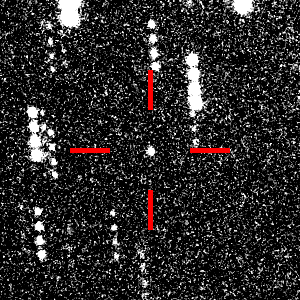}
    \includegraphics[width=6cm]{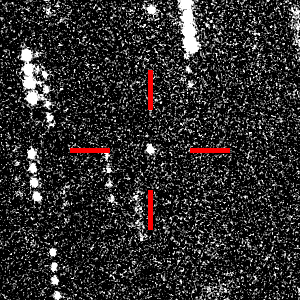}
    \caption{The ``re-scaled mean" combined images used to detect 2023~DZ$_2$ by the STU algorithm. Three subsets of four images each were stacked for detecting this new object.}
    \label{fig:discovery}
    \end{center}
\end{figure*}

The detection was found in CCD3 (Fig.~\ref{fig:discovery}) of the Wide Field Camera (WFC) mounted at the prime focus of the INT, an optical mosaic imaging instrument made of four charge-coupled devices (CCDs). They cover 0.27 square degrees arranged in a 34$\times$34 arcmin$^2$ L-shape design with about 17--22 arcsec gaps between CCDs. The object was identified at an apparent magnitude of 20.3 (in the $G$ band), and it was validated in all stacked images (using the mean, median, trimmed mean, and masked trimmed mean methods) generated as output \citep{stuacm2013proceedings}.  

For the new detection corresponding to 2023~DZ$_2$, STU reported an average rate of apparent motion on sky of 0.7326~arcsec~min$^{-1}$ ($-$0.777~arcsec~min$^{-1}$ in RA, 0.066~arcsec~min$^{-1}$ in DEC, on coordinate motion) for an average position angle of 274{\fdg}6. The object was also independently detected using the Tycho tracker software by one of our ParaSOL collaborators. In order to confirm it, we performed a final check of the images using the Astrometrica software \footnote{\url{http://www.astrometrica.at/}} (the object was detectable in the single exposures). The future releases of our STU pipeline will focus on achieving and proving the real-time candidate validation and reporting capabilities of the software developed within the ParaSOL project. Although in this case only 12 images were obtained during each observing night, the synthetic tracking algorithm allowed us to detect the object automatically in near-real time.

\section{Dynamics\label{sec:dynamics}}
\begin{figure*}
      \centering
         \includegraphics[width=0.332\linewidth]{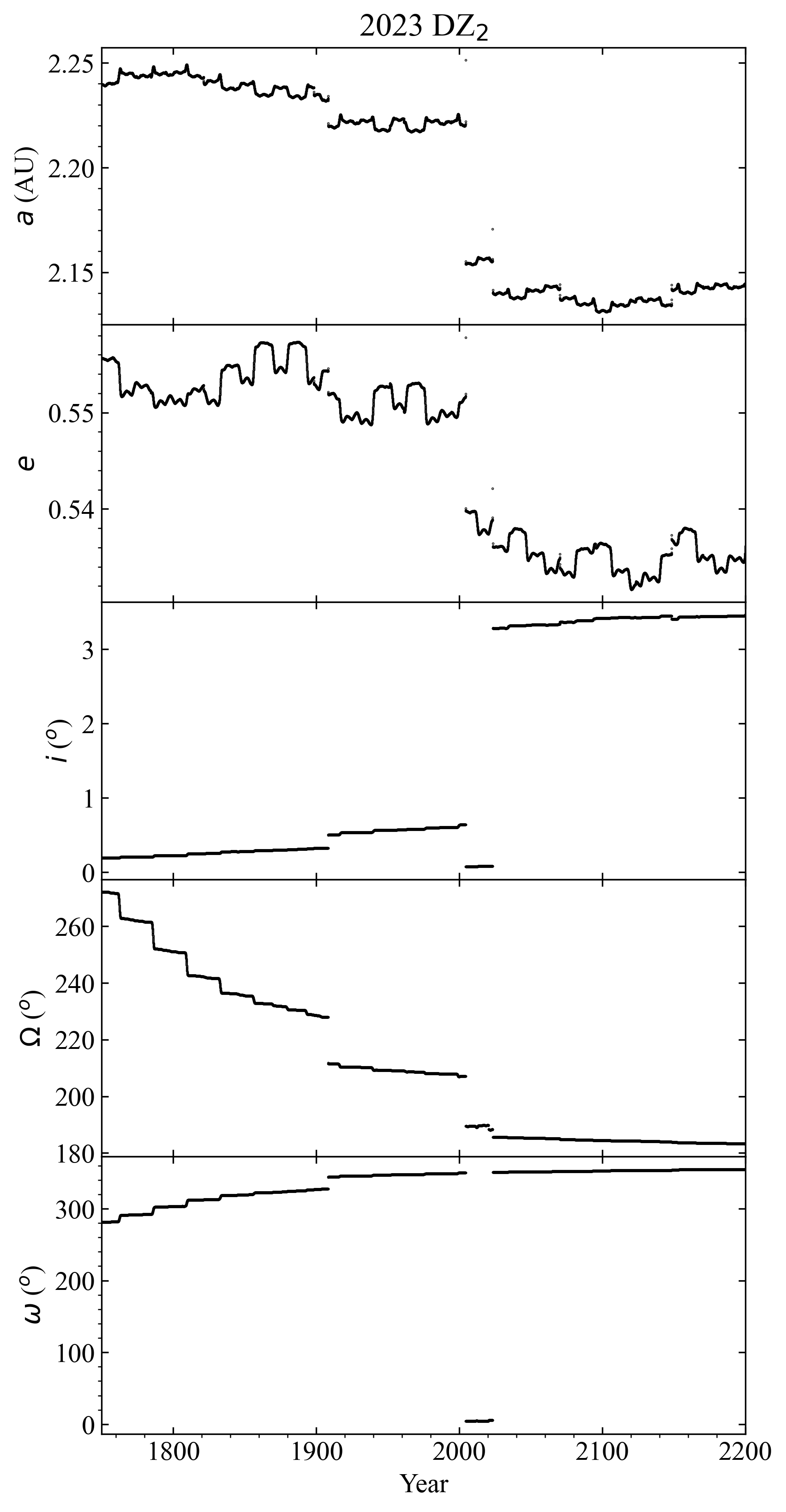}
         \includegraphics[width=0.329\linewidth]{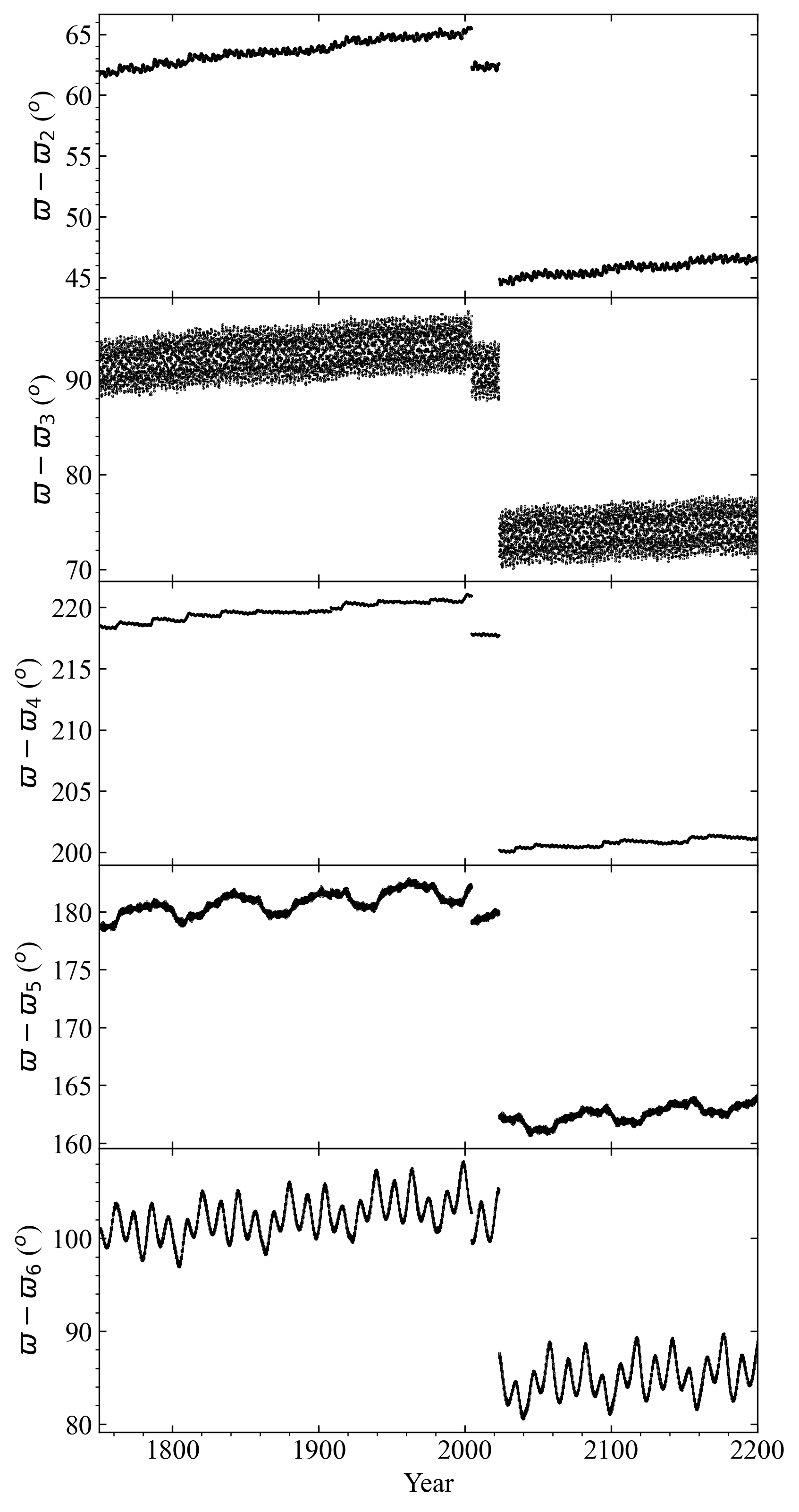}
         \includegraphics[width=0.329\linewidth]{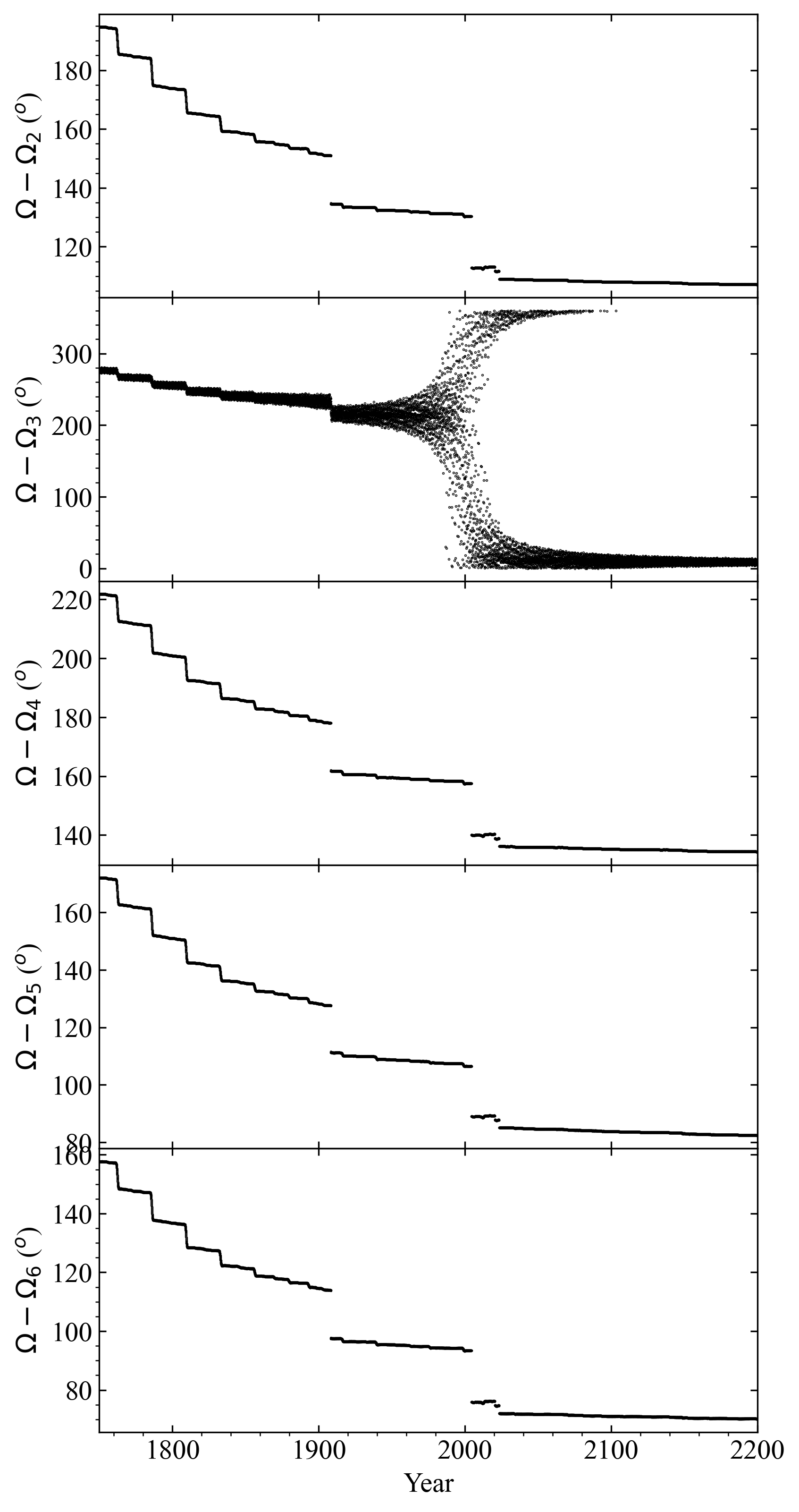}
         \caption{Relevant dynamical evolution of 2023~DZ$_{2}$. \textit{Left panels:} Short-term orbital 
                  evolution of 2023~DZ$_{2}$: $a$, semimajor axis (top panel), $e$, eccentricity (second to top panel), 
                  $i$, inclination (central panel), $\Omega$, longitude of the ascending node (second panel from 
                  bottom), and $\omega$, argument of perihelion (bottom panel). \textit{Central panels:} longitude of 
                  perihelion of 2023~DZ$_{2}$, $\varpi=\Omega+\omega$, relative to that of Venus ($\varpi_{2}$, top 
                  panel), Earth ($\varpi_{3}$,second to top panel), Mars ($\varpi_{4}$, central panel), Jupiter 
                  ($\varpi_{5}$, second panel from bottom), and Saturn ($\varpi_{6}$, bottom panel). An apsidal secular 
                  resonance leads to the libration of the angle $\varpi-\varpi_{i}$ about a constant value (0{\degr} or 
                  180{\degr}). \textit{Right panels:} Longitude of the ascending node relative to that of Venus 
                  ($\Omega_{2}$, top panel), Earth ($\Omega_{3}$,second to top panel), Mars ($\Omega_{4}$, central 
                  panel), Jupiter ($\Omega_{5}$, second panel from bottom), and Saturn ($\Omega_{6}$, bottom panel). A 
                  nodal secular resonance occurs when the angle $\Omega-\Omega_{i}$ librates about a constant value. The 
                  evolution shown here is based on the nominal orbit in Table~\ref{elements} and the output cadence is 15~d. Data source: JPL's {\tt Horizons}.
                 }
         \label{2023dz2dyn}
    \end{figure*}

   Asteroid 2023~DZ$_{2}$ has a Minimum Orbit Intersection Distance (MOID) with Earth of 0.00005~au, one
   of the lowest known among NEAs. This means that the intersection point between its orbit and the
   invariable plane is very close to the path of the Earth--Moon system and that encounters at 1.17~Earth 
   radii are theoretically possible in absence of protective mechanisms such as mean-motion or secular
   resonances. The nominal evolution of the orbit of 2023~DZ$_{2}$ as predicted by JPL's {\tt Horizons} online Solar System data and ephemeris computation 
   service,\footnote{\href{https://ssd.jpl.nasa.gov/horizons/}{https://ssd.jpl.nasa.gov/horizons/}}
   is shown in Fig.~\ref{2023dz2dyn}, left panels. The orbital evolution shows multiple discontinuities
   linked to past and future close encounters of 2023~DZ$_{2}$ with the Earth--Moon system.
   
\begin{figure}
        \centering
        \includegraphics[width=\linewidth]{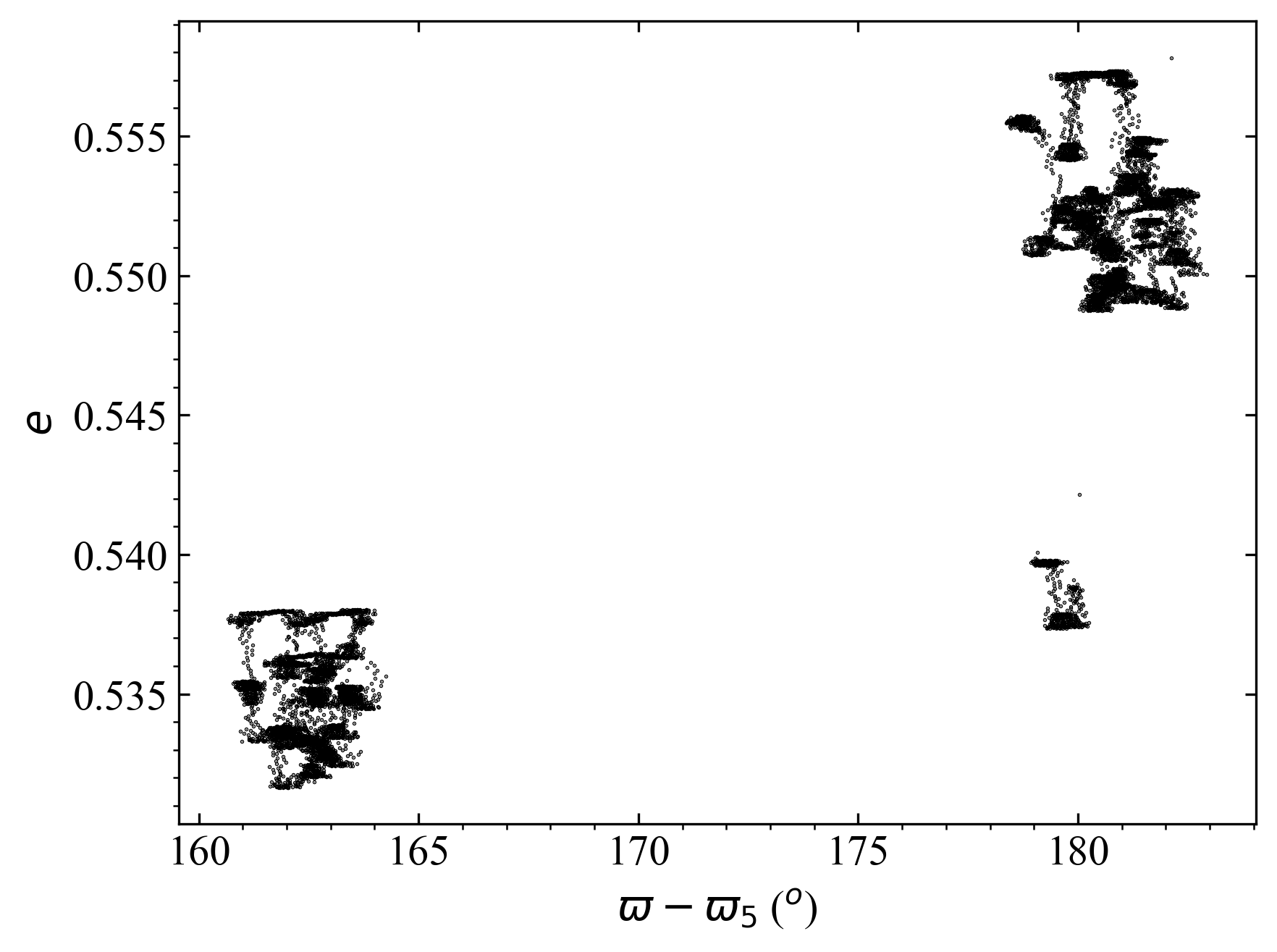}
        \caption{Eccentricity versus $\varpi-\varpi_{5}$ for 2023~DZ$_{2}$.
                }
        \label{nu5evolution}
     \end{figure}

   Table~\ref{elements} shows that the perihelion of 2023~DZ$_{2}$ takes place in close proximity to 
   Earth's orbit, while its aphelion occurs close to the core region of the main asteroid belt. Its 
   semimajor axis at 2.15~au is in the region associated with the so-called $\nu$6 secular resonance in
   which the value of the longitude of perihelion ($\varpi=\Omega+\omega$, see e.g. 
   \citealt{1999ssd..book.....M}) of a minor body relative to
   that of Saturn oscillates about 0{\degr} or 180{\degr} (see e.g. \citealt{1989CeMDA..46..231F,1991CeMDA..51..169M}). 
   This is the secular apsidal resonance with Saturn. Figure~\ref{2023dz2dyn}, central panels shows the nominal evolution of the critical angle associated with the apsidal resonances of 2023~DZ$_{2}$ with Venus, Earth, Mars, Mercury, Jupiter, and Saturn. While 2023~DZ$_{2}$ is not currently subjected to the
   $\nu$6 secular resonance, it appears to have been subjected to the $\nu$5 secular resonance with Jupiter in the recent past. As
   the associated critical angle librated about 180{\degr} (see Fig.~\ref{nu5evolution}), 2023~DZ$_{2}$ reached aphelion when Jupiter was at perihelion. This explains why the MOID of 2023~DZ$_{2}$ with Jupiter amounts to 1.636~au.
   On the other hand, Fig.~\ref{2023dz2dyn}, right panels, reveals that 2023~DZ$_{2}$ is in nearly nodal
   resonance with Earth as the ascending node of this small body is in the path of our planet. 
   
   In summary, the dynamics of 2023~DZ$_{2}$ is controlled by Earth and Jupiter, with Earth currently being a direct perturber as a result of the near nodal resonance and Jupiter playing the role of secular perturber that projected a near apsidal resonance on 2023~DZ$_{2}$. The fact that this object is not subjected to the $\nu$6 secular resonance could explain why its MOID with Earth is so small yet its probability of impact in the near future is zero. We may argue that NEA 2023~DZ$_{2}$ was somewhat protected against collision with Earth by a near $\nu$5 secular resonance. The data discussed above were retrieved from JPL's Small-Body Database (SBDB),\footnote{\href{https://ssd.jpl.nasa.gov/tools/sbdb\_lookup.html\#/}
   {https://ssd.jpl.nasa.gov/tools/sbdb\_lookup.html\#/}} which is provided by the Solar System Dynamics Group (SSDG, \citealt{2011jsrs.conf...87G,2015IAUGA..2256293G}),\footnote{\href{https://ssd.jpl.nasa.gov/}{https://ssd.jpl.nasa.gov/}} and {\tt Horizons} using tools provided by the {\tt Python} package {\tt Astroquery} \citep{2019AJ....157...98G} and its {\tt HorizonsClass} class.\footnote{\href{https://astroquery.readthedocs.io/en/latest/jplhorizons/jplhorizons.html}{https://astroquery.readthedocs.io/en/latest/jplhorizons/jplhorizons.html}}

%
%-----------------------------------------------------------
%
%
%-----------------------------------------------------------
%
%-------------------------------------------------------------------------------------------------------------------------------------------
%
%
%-------------------------------------------------------------------------------------------------------------------------------------------
%

   The analysis above is telling us nothing about the role of the uncertainties on the computed evolution
   of this NEA but relatively large uncertainties coupled with recurrent encounters at close range may 
   severely hamper our ability to explore the orbital evolution of this object beyond a few decades from 
   the current epoch (see e.g. \citealt{2003A&A...408.1179V}). In order to investigate the role of the 
   uncertainties on the reconstruction of the past evolution of this NEA and on the prediction of its 
   future behaviour, we have performed $N$-body simulations using a direct $N$-body code developed by 
   \citet{2003gnbs.book.....A} that implements the  Hermite integration scheme formulated by 
   \citet{1991ApJ...369..200M}. The code is publicly available from the website of the Institute of 
   Astronomy of the University of Cambridge.\footnote{\href{http://www.ast.cam.ac.uk/~sverre/web/pages/nbody.htm}{http://www.ast.cam.ac.uk/~sverre/web/pages/nbody.htm}} Relevant results from this code were discussed in detail by \citet{2012MNRAS.427..728D} that also includes many technical details. 

\begin{figure}
        \centering
        \includegraphics[width=\linewidth]{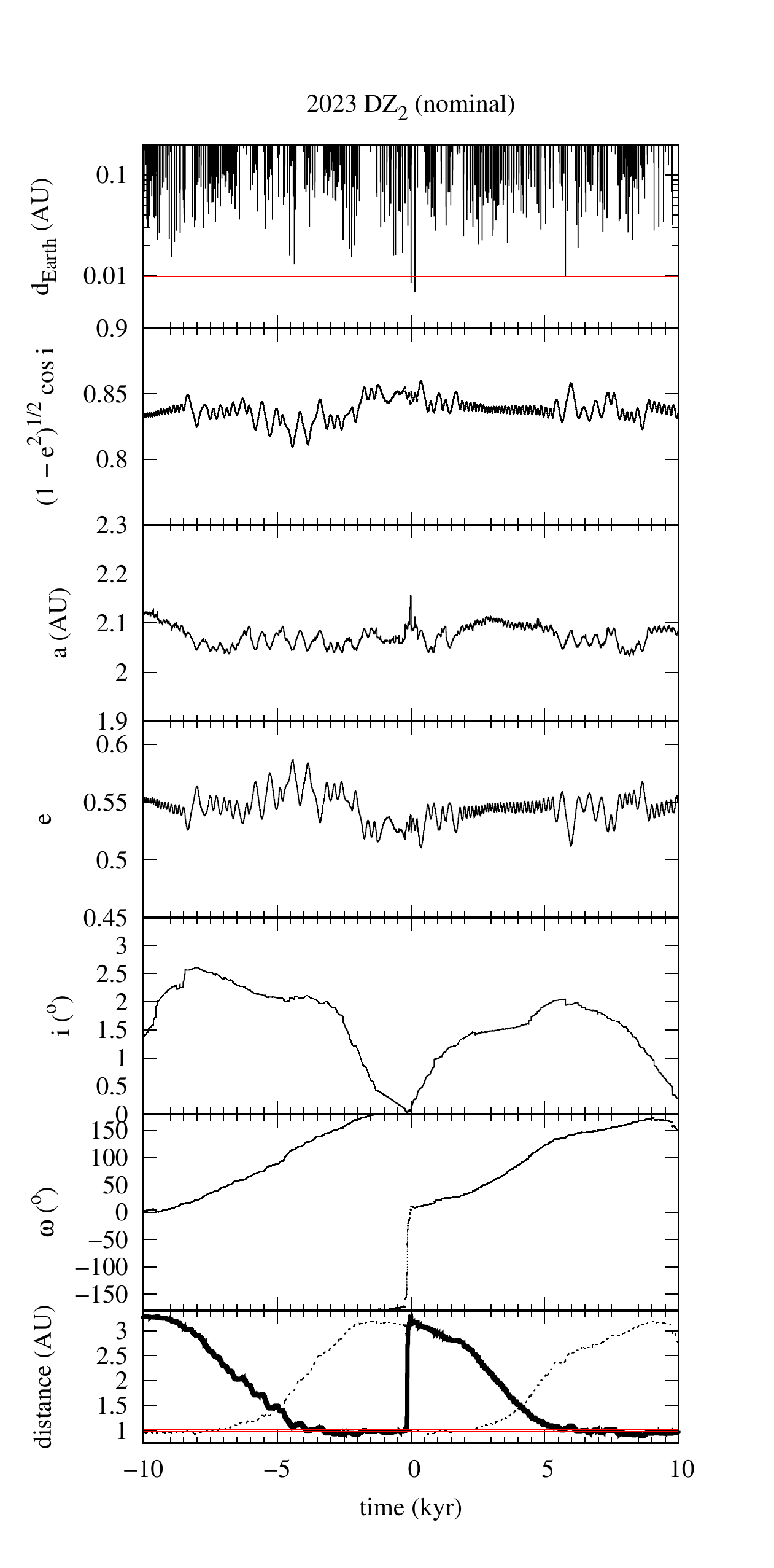}
        \caption{Evolution over time of the values of the orbital elements and other relevant parameters 
                 for the nominal orbit of 
                 2023~DZ$_{2}$ as in Table \ref{elements}. The top panel shows the evolution of the geocentric distance; the value of the Hill radius of the Earth, 
                 0.0098~au, is plotted as reference. The second to top panel focuses on the evolution of the value of the Kozai--Lidov parameter. The following four panels
                 show respectively the 
                 evolution of the values of semimajor axis, eccentricity, inclination, and argument of perihelion of the nominal orbit. the panel at the bottom displays the
                 distances from the descending (thick line) and ascending nodes (dotted line) to the Sun; 
                 Earth's aphelion and perihelion distances are shown as well in red. The output time-step size is 0.1~yr. The source of the input data is JPL's SBDB. The standard epoch JD 2460000.5 (25-Feb-2023) TDB that is the origin of times.
                }
        \label{nominal}
     \end{figure}

\begin{figure}
        \centering
        \includegraphics[width=\linewidth]{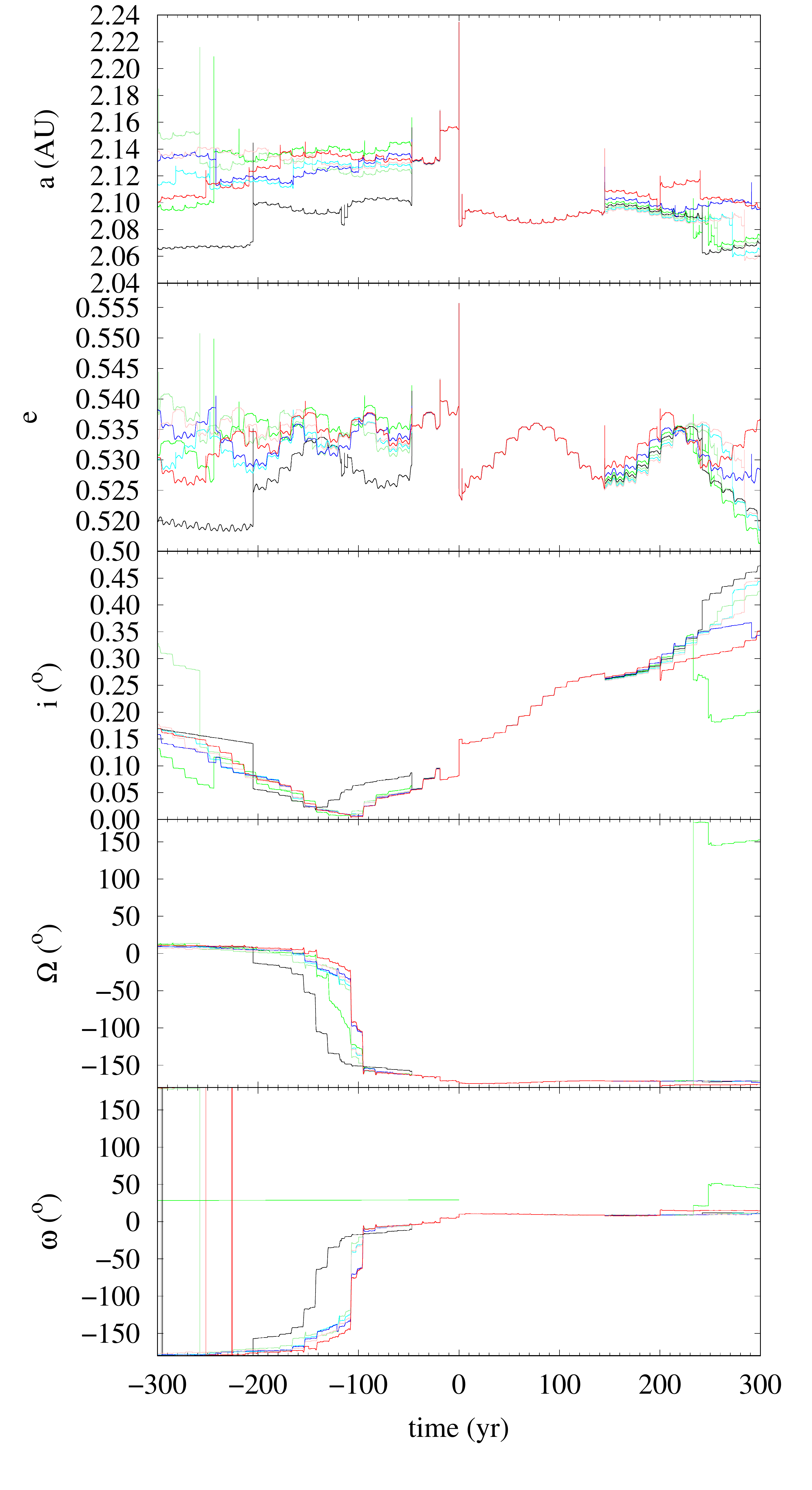}
        \caption{Evolution of the values of the semimajor axis ($a$, top panel), eccentricity ($e$, second to top panel), 
         inclination ($i$, third to bottom panel), ascending node ($\Omega$, second to bottom panel), and argument of perihelion ($\omega$, bottom panel) of 2023~DZ$_{2}$. The panels display results of integrations for the 
         nominal orbit (in black) and those of control orbits with Cartesian vectors separated $\pm$3$\sigma$ (in light green, $-$3$\sigma$, and green, 3$\sigma$), 
         $\pm$6$\sigma$ (in cyan, $-$6$\sigma$, and blue, 6$\sigma$), and
         $\pm$9$\sigma$ (in pink, $-$9$\sigma$, and red, 9$\sigma$).
         The output time-step size is 4.383~h. The 
         source of the input data is JPL's SBDB and they are referred to the standard epoch JD 2460000.5 (25-Feb-2023) TDB that is the origin of times.
                }
        \label{uncertain}
     \end{figure}

   Figure~\ref{nominal} shows our results for the orbital evolution of the nominal orbit of 2023~DZ$_{2}$
   in the time interval ($-$10,000, 10,000)~yr around the standard epoch JD 2460000.5 (25-Feb-2023) TDB
   that is the origin of times in the figure. The enhanced probability of encounters at close range is
   connected to the fact that the nodal distances experience an oscillation that periodically places the 
   nodes of 2023~DZ$_{2}$ in the path of Earth (Fig.~\ref{nominal}, bottom panel). On the other hand, the value of the Kozai--Lidov parameter (Fig.~\ref{nominal}, second to top panel) does not experience
   obvious oscillations and we can discard that 2023~DZ$_{2}$ could be subjected to a von Zeipel-Lidov-Kozai secular resonance\citep{1910AN....183..345V,1962P&SS....9..719L,1962AJ.....67..591K,2019MEEP....7....1I}. 
   Figure~\ref{uncertain} shows the short-term evolution of control orbits with initial conditions separated from those of the nominal orbit. From the calculations, we find that predicting the future evolution of this NEA becomes difficult after 12-April-2165, when 2023~DZ$_{2}$ will experience another encounter with Earth,
   this time at 0.01~au. Investigating its orbital past is also challenging because a relatively distant
   encounter with Jupiter on 17-March-1976 placed the object in its present trajectory.
   Between these two dates, all the control orbits up to $\pm$9$\sigma$ from the nominal solution provide
   a consistent picture of the orbital evolution of this NEA.
%
%-------------------------------------------------------------------------------------------------------------------------------------------
%
%
%-------------------------------------------------------------------------------------------------------------------------------------------
%
%
%-------------------------------------------------------------------------------------------------------------------------------------------
%
%
%-------------------------------------------------------------------------------------------------------------------------------------------
%

   So far we have included the uncertainties by assuming that they are uncorrelated. This is a valid assumption when the orbit
   determination is robust, based on a large number of high-quality observations spanning a long time interval. This is however
   not the case for 2023~DZ$_{2}$. In order to account for any correlations present in the data, we carried out additional
   integrations backward and forward in time using initial conditions corresponding to control or clone orbits produced by the 
   Monte Carlo using the Covariance Matrix (MCCM) approach described by \citet{2015MNRAS.453.1288D}. These synthetic orbits are 
   based on the nominal orbit determination (see Table~\ref{elements}) with random noise added on each orbital element by making 
   use of the covariance matrix. The covariance matrix was retrieved from JPL's SSDG SBDB using the {\tt Python} package 
   {\tt Astroquery} and its {\tt SBDBClass}\footnote{\href{https://astroquery.readthedocs.io/en/latest/jplsbdb/jplsbdb.html}{https://astroquery.readthedocs.io/en/latest/jplsbdb/jplsbdb.html}}  
   class, and it is referred to epoch 2460023.5 (20-Mar-2023) TDB that is the origin of times for the new calculations. The MCCM
   methodology was used to generate initial positions and velocities for 10$^{3}$ control orbits that were evolved dynamically 
   using the direct $N$-body code.
%
%-------------------------------------------------------------------------------------------------------------------------------------------
%
     \begin{figure}
        \centering
        \includegraphics[width=\linewidth]{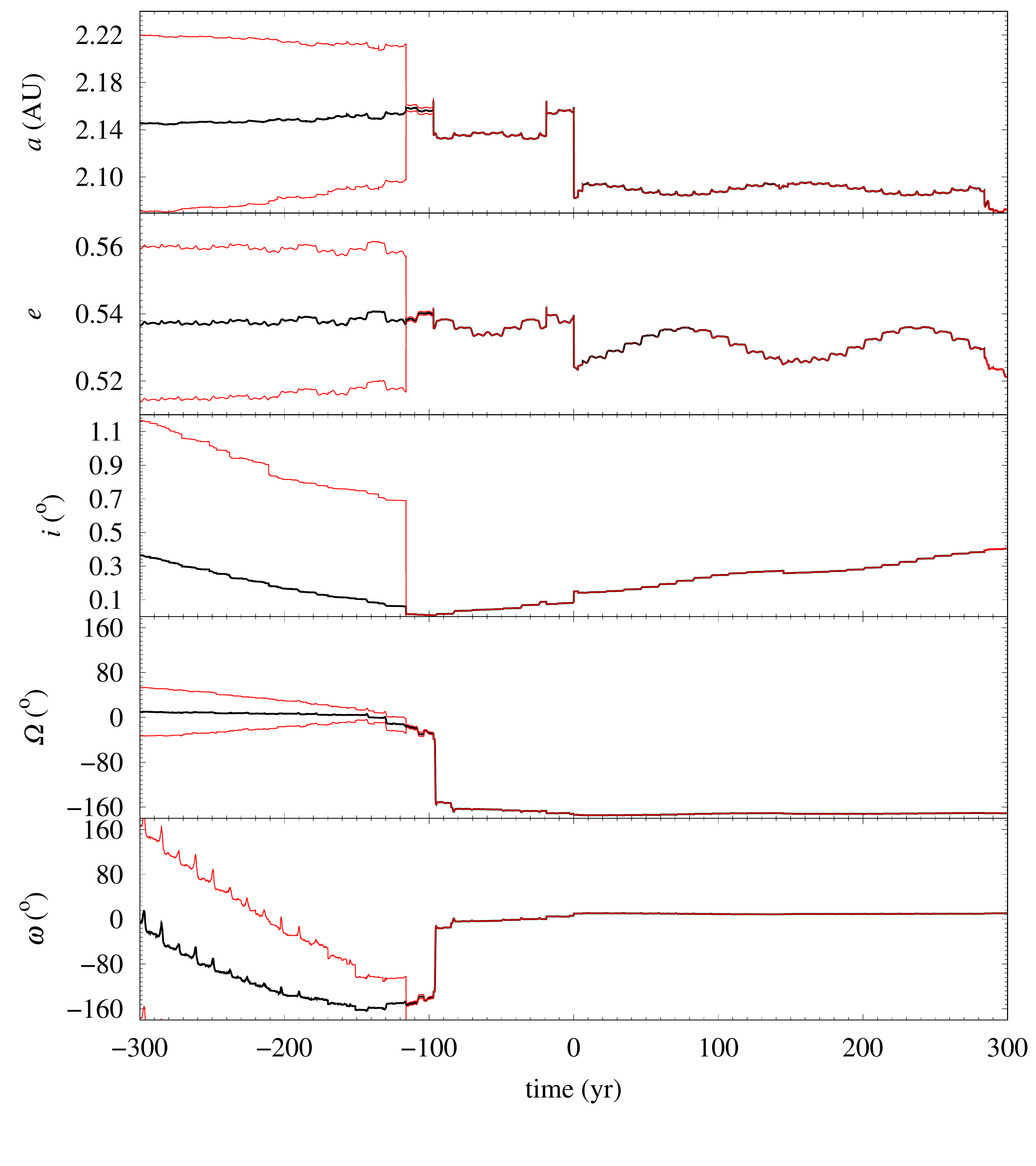}
        \caption{Evolution of the values of the semimajor axis ($a$, top panel), eccentricity ($e$, second to top panel), 
         inclination ($i$, third to bottom panel), ascending node ($\Omega$, second to bottom panel), and argument of perihelion ($\omega$, bottom panel) of 2023~DZ$_{2}$ according to the MCCM approach. The panels display results of the integrations of 10$^{3}$ control orbits with 
         initial positions and velocities generated using the MCCM methodology. In black, we display 
         the average evolution of the orbital element and in red we show the range linked to the 1-$\sigma$ uncertainty or the 16th and 84th percentiles. The output time-step size is 0.1~yr. The source of the input data is JPL's SBDB and they are referred to epoch 2460023.5 (20-Mar-2023) TDB that is the origin of times.
                }
        \label{cova2023DZ2}
     \end{figure}
%
%-------------------------------------------------------------------------------------------------------------------------------------------
%

   Figure~\ref{cova2023DZ2} shows the result of the past and future evolution of 2023~DZ$_{2}$ according to the MCCM approach. In
   black, we display the average evolution and in red we show the range linked to the 1-$\sigma$ uncertainty or the 16th and 84th 
   percentiles. While the uncertainty in the reconstruction of the past orbital evolution of this NEA is large beyond 100~yr into
   the past, the one associated with predictions of its future orbital behaviour is far smaller. This is the result of particularly
   close encounters with the Earth--Moon system when the orbital inclination of 2023~DZ$_{2}$ became virtually zero, over one 
   century ago. Therefore, its current orbit determination is not robust enough to investigate its origin and how this object was
   inserted in NEA orbital parameter space. More observations are needed to produce a better orbit determination for such a study. 
   On the other hand, our results based on the covariance matrix are quite consistent with those obtained assuming uncorrelated 
   uncertainties.

\section{Time-series photometry\label{sec:photom}}
\label{lightcurve}

Photometric data were obtained during three consecutive nights at the "Telescopios Gemelos de Dos Metros" (Two Meter Twin Telescopes --- TTT) facility. This is located located at the Teide Observatory (latitude: 28{\degr}~18'~01.8"~N; longitude: +16{\degr}~30'~39.2"~W, and an altitude of 2386.75~m), in the island of Tenerife (Canary Islands, Spain). Currently, it includes two telescopes (called TTT1 and TTT2) with an aperture of 0.80 m, installed on altazimuth mounts, and with focal ratios of f/4.4 and f/6.8, respectively.

The observations were made using the QHY411M\footnote{\url{https://www.qhyccd.com/}}  cameras \citep{2023PASP..135e5001A} installed on one of the Nasmyth ports of each telescope. They are equipped with sCMOS (scientific Complementary Metal–Oxide–Semiconductor) image sensors consisting in 151 megapixels with a pixel size of 3.76~$\mu$m~pixel$^{-1}$. This provides an effective FoV of 51.4$^{\prime}\times$38.3$^{\prime}$ (with an angular resolution of 0.22"~pixel$^{-1}$) in TTT1 and 33.1'$\times$24.7' (angular resolution of 0.14"~pixel$^{-1}$) in TTT2. In all the observing runs a band-pass filter (the band covers the 0.4 to 0.7 $\mu$m wavelength interval) was used. The exposure time was dynamically set between 10 and 20~s to ensure a signal to noise ratio (SNR) larger than 50. A total of five observing runs were performed, one with TTT2 on March 20 and two simultaneous runs with each telescope on the nights of March 21 and 22. Table \ref{tab:obs} shows the effective coverage time for each night.

%Photometric data were obtained on three consecutive nights with the TTT1 and TTT2 telescopes , located at the Teide Observatory, in the island of Tenerife (Canary Islands, Spain). These are two 0.80-m Alt-Az telescopes f/4.4 and f/6.8 respectively, which are currently under commissioning. The observations were made using the QHY411M cameras \citep{Alarcon2023} installed on one of the Nasmyth ports of each telescope. They are equipped with 151 MPx 3.76 micron sCMOS sensors, resulting in an effective FOV of 51.4$\times$38.3 0.22"/px in TTT1 and 33.1$\times$24.7 0.14"/px in TTT2. In all the observing runs a UV/IR-Cut CMOS-optimized filter, almost equivalent to SDSS g$^\prime+$r$^\prime$, was used. The exposure time was dynamically set between 10 and 20 s to ensure a SNR$>$50. A total of 5 observing runs were performed, one with TTT2 on March 20 and two simultaneous runs with each telescope on the nights of March 21 and 22. Table \ref{tab:obs} shows the effective coverage time for each night.

The images were bias and sky flat-field corrected. Then, they were trimmed and binned 2$\times$2. Aperture photometry was performed using Tycho Tracker software. The images were aligned with bicubic interpolation and down sampled by a factor 2 for astrometric calibration which was performed with Astrometry.net \citep{Lang2010}. 

To obtain the photometry of the object, a fixed aperture of 2$\times$FWHM (Full Width at Half Maximum of the point spread function of stars) in the first image of each set was used. An outer ring with the inner radius located at 4$\times$FWHM was used to estimate the sky background signal. The same apertures were used for the comparison stars, selected constraining $0.60<(B-V)<0.70$. The initial and final positions of the asteroid were marked manually in the images in order to prevent any miss-identification by the algorithm.

\begin{figure}[h!]
    \begin{center}
    \includegraphics[width=1\columnwidth]{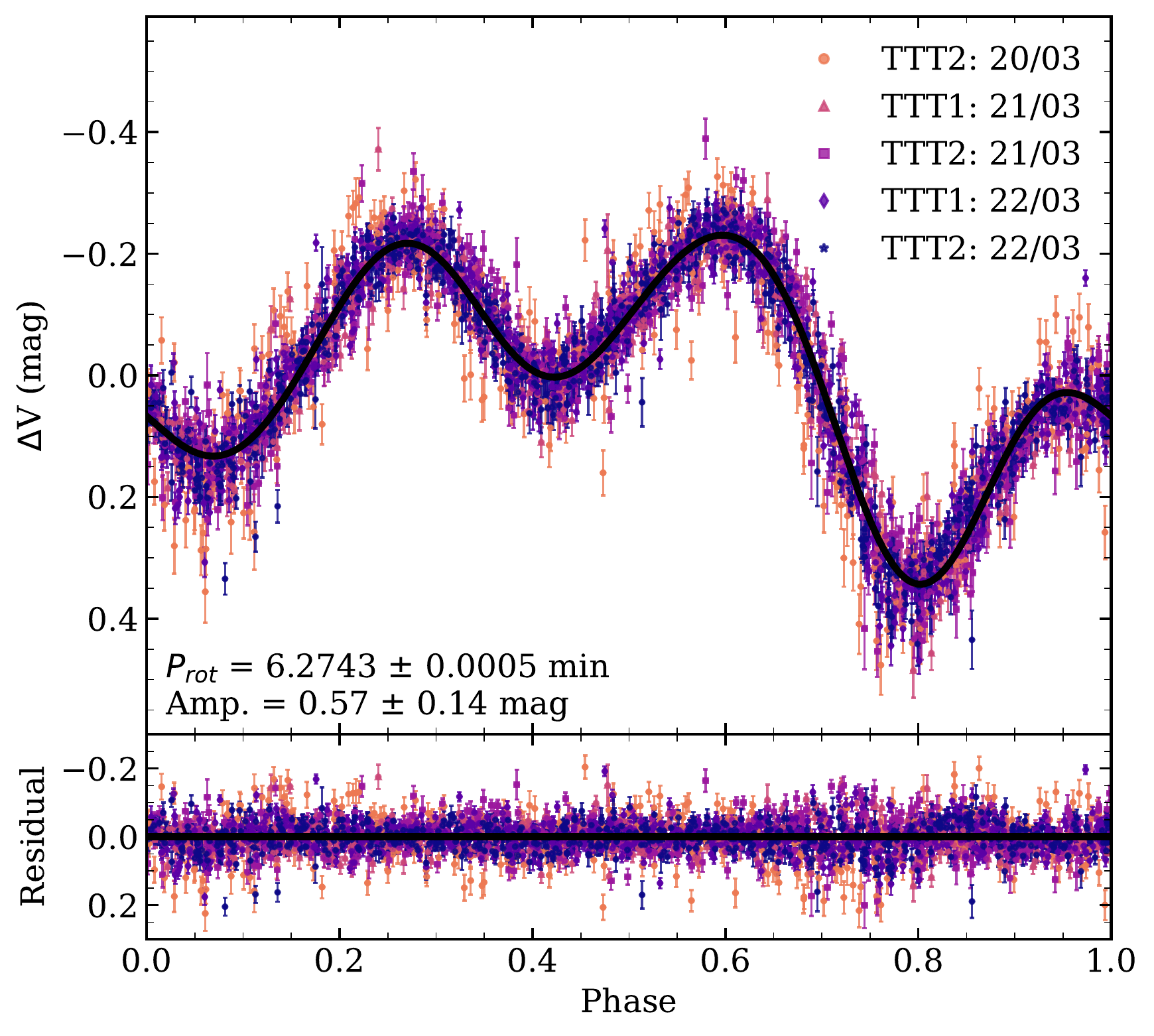}
    \caption{Phased light curve of 2023~DZ$_{2}$ computed from photometric measurements obtained by the TTT1-2 telescopes. The rotation period and amplitude of the curve are shown at the lower left corner of the top panel. The total coverage is 9.8~h, distributed over three consecutive nights. Residuals are shown in the bottom panel.}
    \label{fig:lc_TTT}
    \end{center}
\end{figure}

Photometric measurements were extracted and corrected for distance and light--time. The five--term Lomb-Scargle periodogram was obtained, with a well-marked peak in the power spectrum centred at a period of $P_{\rm rot}=6.2743\pm0.0005$~min and its aliases. As uncertainty, 1--$\sigma$ of the Gaussian curve fitted to the exponentiated power peak was taken \citep{VanderPlas2018}. The phased light curve is shown in Figure \ref{fig:lc_TTT}. The amplitude of the curve obtained, considering photometric errors, is $0.57\pm0.14$~mag.

The initial observations performed with TTT1 and TTT2 show a super-fast rotator asteroid \citep{2023MNRAS.tmp..693L}. The $P_{\rm rot}=6.274$ min is indicative of intrinsic strength to resist centrifugal disruption, otherwise 2023~DZ$_{2}$  would break apart \citep{2000Icar..148...12P}. It could be a coherent body or monolith \citep[e.g.][]{2020MNRAS.495.3990M, 2014M&PS...49..788S}. The data existing in Asteroid Lightcurve Database (LCDB, \citealt{2009Icar..202..134W}),\footnote{\url{https://minplanobs.org/mpinfo/php/lcdb.php}} updated on February 2023 shows that no faster rotator correspond to a low (i.e. $\leq0.10$) albedo asteroid \citep{2023MNRAS.521.3784L}. Thus, at first glance, the period determination adds a strong constraint on the cohesive strength of the asteroid and it is a strong indicative of a high albedo asteroid.

The hypothesis of a high albedo object is supported by the spin limit for small asteroids \citep{2020MNRAS.495.3990M,2020PASP..132f5001R}. At an absolute magnitude $H\approx24$~mag and an albedo of $p_V\leq0.1$, this asteroid will have a size larger than 60~m which for a rotation period of $P_{\rm rot}=6.2743\pm0.0005$~min is outside the spin barrier determined by \citep{2020PASP..132f5001R} for C-type asteroids. 

The three peaks of the light curve show an irregular shape which favours the hypothesis of a monolithic body. This result was confirmed by observations performed on March 25 with the Goldstone Radar. The preliminary result reported on their webpage\footnote{\url{https://echo.jpl.nasa.gov/asteroids/2023DZ2/2023DZ2.2023.goldstone.planning.html}} shows that ``at some orientations the shape looks somewhat triangular; at others it looks rounded; there is a flat area, and a small-scale topographic feature that appears on the leading edge".

\section{Spectro-photometry and spectroscopy}

The spectro-photometric (photometric observations performed with various broad-band filters) and the spectral observations represent the best ground-based observing techniques to constrain the composition properties of small objects. The spectro-photometric method is more suitable for characterizing fainter objects since the incoming flux is collected in each bandpass and not dispersed as in the case of spectral observations. The faintest targets, which cannot be spectroscopically observed with any other telescope can, be taxonomically classified using this technique. The disadvantage is the impossibility of making a distinction between different sub-types of the same taxonomic complex. This is because a colour is just the normalized difference in the amount of reflected light-flux between the wavelength ranges covered by a broad band filter, while the specific features may cover a short spectral interval. Thus, the spectro-photometry is the first approximation of the spectral properties. %Hence, the spectral observations are required for a precise classification.

We were able to acquire spectro-photometric data for 2023 DZ$_2$ on the night 22-March-2023 using the Telescopio Carlos S\'{a}nchez (TCS) from Teide Observatory. Although these observations were performed after we acquired spectra using the GTC telescope it is worth to mention them as the first approach to infer the taxonomic classification, in order to highlight the strength of this technique for fainter asteroids. 

The observations were performed with the MuSCAT2 (the acronym is derived from ``a multicolour simultaneous camera for studying atmospheres of transiting exoplanets") imaging instrument \citep{2019JATIS...5a5001N} which is mounted on the Cassegrain focus of the TCS. This configuration allows us to perform simultaneous photometric observations in four visible broad-band filters, namely \textit{g}\,(400\,-\,550), \textit{r}\,(550\,-\,700), \textit{i}\,(700\,-\,820) and \textit{z$_s$}\,(820\,-\,920)\,nm. At the end of each of the four channels there is an independently controllable CCD camera (1024$\times$1024 pixels), having a pixel size of $\sim$0.44"~pixel$^{-1}$ and a FoV of $7.4^{\prime}~\times~7.4^{\prime}$.

\begin{figure}[h!]
    \begin{center}
    \includegraphics[width=1\columnwidth]{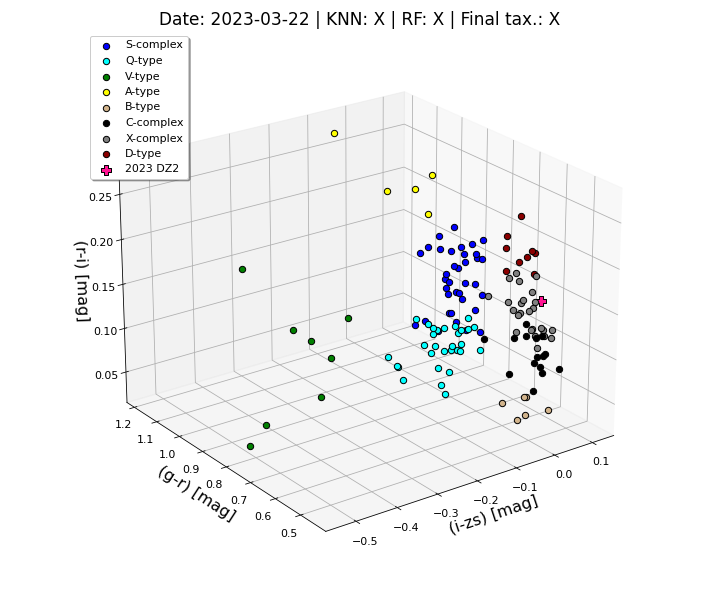}
    \caption{The (g-r) vs. (r-i) vs. (i-z$_s$) colour-colour diagrams of the 155 objects with known spectral classification, used to classify 2023~DZ$_{2}$ based on the TCS/MusSCAT2 data. The taxonomic types defined in \citet{2009Icar..202..160D} system have been divided in three major composition groups, namely the Q\,/\,S-complex (green and blue dots), C-complex (black dots) and X-complex (grey dots). Besides them, three end-member types are considered, A-, D- and V-type.}
    \label{fig:DZ2colors}
    \end{center}
\end{figure}

In order to obtain the light-curves and the colours, we used the PHOTOMETRYPIPELINE -- PP \citep{2017A&C....18...47M}. This is a software package written in Python which obtains calibrated photometry from the FITS images by performing the astrometric registration, aperture photometry, photometric calibration, and asteroid identification. PP is written in Python and uses the \emph{Astromatic} suite\footnote{\url{https://www.astromatic.net/}}, namely SExtractor 
for source identification and aperture photometry \citep{1996A&AS..117..393B}, SCAMP for astrometric calibration \citep{2006ASPC..351..112B}, and SWarp for image re-gridding and co-addition \citep{2002ASPC..281..228B}. It also uses JPL's Horizons module for obtaining the SSOs ephemeris in order to identify them in the images.

For astrometric registration, we used the GAIA DR2 catalog \citep{2018A&A...616A...1G}.  In order to improve the accuracy, we applied the registration algorithm twice for each image. We discarded all those images for which the astrometric registration failed. The Pan-STARRS catalog \citep{2012ApJ...750...99T} was used for all photometric calibrations. The accuracy of these calibrations is dependent on the number of stars imaged by each exposure (19 to 40 stars were used for the photometric calibration of each image). 

PP uses aperture photometry performed by SExtractor. We applied the PP algorithm that finds the optimum aperture radius based on a curve-of-growth analysis \citep{2000hccd.book.....H, 2017A&C....18...47M}. From each set of four simultaneous images, we computed the $(g-r)$, $(r-i)$, and $(i-z_s)$ colours. The reported values of the colours represent the median of data obtained during four hours of observations. 

We used these colour values to infer the taxonomic classification. Thiswas carried out in a robust way with the K-Nearest neighbours (KNN) and Random Forest (RF) algorithms. These were implemented using the {\tt Python} package {\tt SCIKIT-LEARN}. The KNN algorithm classifies an object based on the label values or taxonomy of its neighbours in the colour-colour diagram, while the RF algorithm assigns the final label to an object using decision-tree structures.

Both algorithms require a training set which includes objects for which we know both the photometric colours (from our data set) and spectral data (from the literature).  To generate it we searched the available spectral information of all the objects with TCS colours in the SMASS-MIT-Hawaii Near-Earth Object Spectroscopic Survey (MITHNEOS MIT-Hawaii Near-Earth Object Spectroscopic Survey) program \citep{2019Icar..324...41B} and the Modeling for Asteroids (M4AST) database \citep{2012A&A...544A.130P}. We retrieved spectral classifications for 84 of the NEAS observed as well by our TCS/MuSCAT2 program \citep{2021EPSC...15..820P}. To increase the training sample, we computed the synthetic colours using  the visible spectra published by \citet{2019A&A...627A.124P} and \citet{2018P&SS..157...82P}. The final training sample consisted of 154 asteroids classified as 5 A-type, 8 V-type , 34 Q-types and 48 S-complex, 7 B-types, 15 C-complex, 9 D-types, and 28 X-complex.

In order to account for magnitude errors determined on each of the broad-band filters, we applied a Monte-Carlo approach. We started from the colour value and its error and generated three normal distributions (one for each colour) of 10\,000 fictitious colour values. Then, for each of these cases we classified the object. Finally, the assigned taxonomy was the one with the highest frequency. Based on the colour values of $(g-r) = 0.555\pm0.055$~mag, $(r-i) = 0.154\pm0.055$~mag, and $(i-z_s) = 0.064\pm0.059$~mag both algorithms classify this object as an X-complex member with 100\% probability (this probability indicates that all the clones generated within the Monte-Carlo approach were classified as X-complex objects).

The most powerful technique used to characterize this object was spectroscopy. The visible spectra of 2023~DZ$_{2}$ were obtained on two separate nights: one on the night of 17-March-2023, and another one on the night of 20-March-2023, using the OSIRIS camera-spectrograph \citep{Cepa2000,Cepa2010} at the 10.4~m GTC, under the program GTC31-23A. The telescope is located at the El Roque de Los Muchachos Observatory, in the island of La Palma (Canary Islands, Spain). 

The OSIRIS instrument (upgraded in January 2023) %, is equipped with a new blue-sensitive monolithic 4K$\times$4K pixels detector that provides a total unvignetted field of view of 7.8$\times$7.8 arcmin$^2$. 
was employed on both nights. We used the 1.2" slit and the R300R grism (resolution $R$=348 for a 0.6" slit, dispersion of 7.74 \AA~pixel$^{-1}$) covering the 0.48--0.92~$\mu$m wavelength range. The slit was oriented along the parallactic angle to minimize the effects of atmospheric differential refraction and the telescope tracking was set at the asteroid proper motion. Details of the observational circumstances are shown in Table~\ref{tab:obs}. Two spectra of 300~s of exposure time each were obtained, with an offset of 10" in the slit direction in between them. To obtain the asteroid reflectance spectrum, we observed two solar analogue stars from the Landolt catalogue \citep{Landolt1992}, SA98-978 and SA102-1081 at a similar airmass as that of the asteroid.  

Data reduction was completed using standard procedures. The images were bias and flat-field corrected. Sky background was subtracted and a one-dimensional spectrum was extracted using a variable aperture, corresponding to the pixel where the intensity was 10\% of the peak value. Wavelength calibration was carried out using Xe+Ne+HgAr lamps. This procedure was applied to the spectra of the asteroid and the stars. We then divided the asteroid's individual spectra by the spectra of the solar analogues, and the resulting ratios were averaged to obtain the final reflectance spectrum of 2023~DZ$_{2}$.

\begin{figure}[h!]
    \begin{center}
    \includegraphics[width=1\columnwidth]{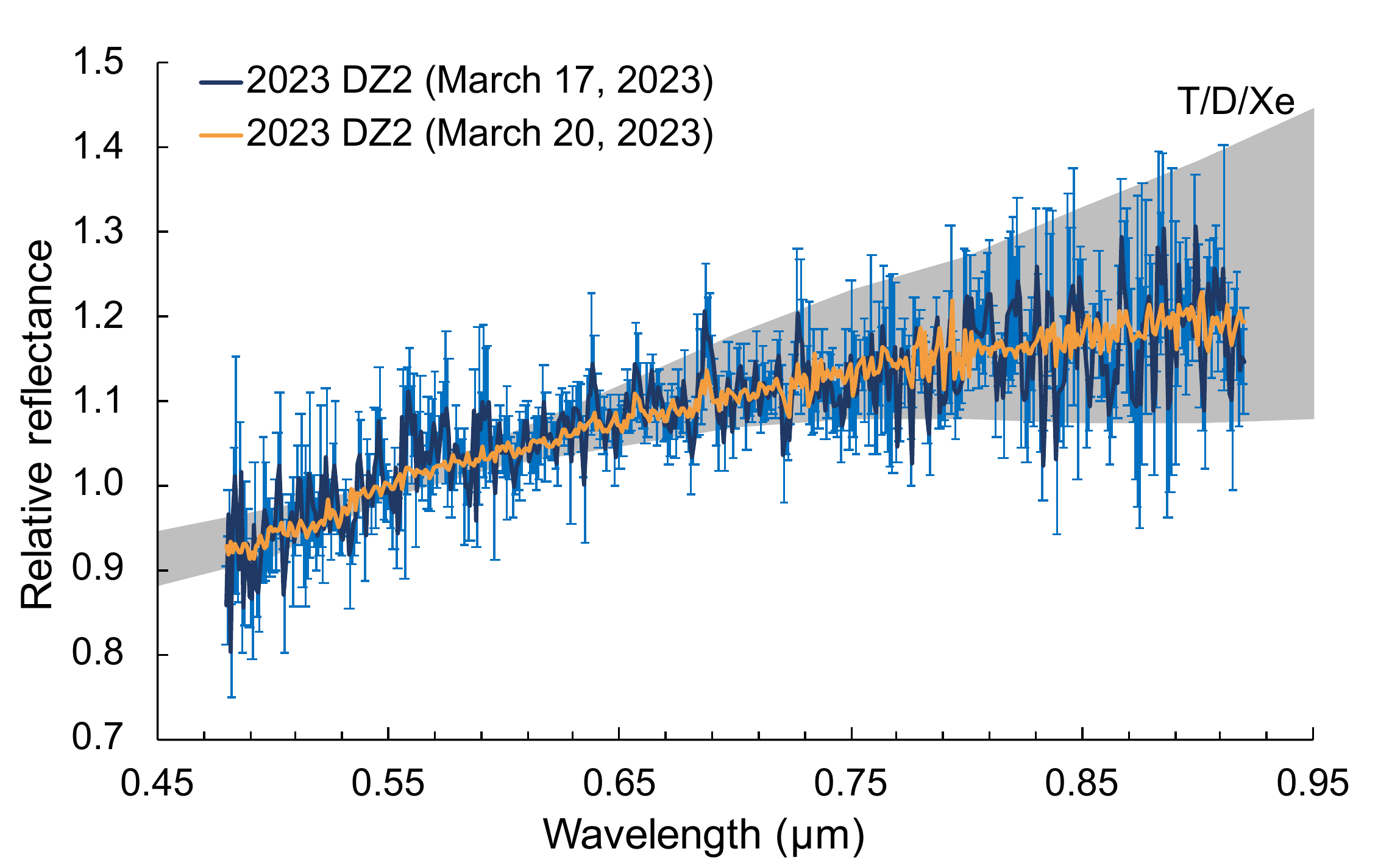}
    \caption{Visible spectra of asteroid 2023~DZ$_{2}$ obtained with the 10.4~m Gran Telescopio Canarias (GTC) on the night of 17-March-2023 (dark blue) and on the night of 20-March-2023 (orange). Error bars correspond to the standard deviation of the mean for the March 17 data. The error bars for the March 20 data are much smaller and are contained within the larger ones. The grey hatched region accounts for the three best taxonomic fits, in order of increasing $\chi^2$: T, D, and Xe-types.}
    \label{fig:DZ2spectrum}
    \end{center}
\end{figure}

The spectrum obtained on the night of 17-March-2023 is shown in Fig.~\ref{fig:DZ2spectrum} in dark blue, together with the error bars associated to the standard deviation of the average. The spectrum obtained on the night of March 20, when the asteroid was one apparent magnitude brighter, is shown in the same figure in orange. The agreement between the two spectra is perfect. Finally, the obtained spectra were used to classify 2023~DZ$_{2}$ taxonomically. We did so using the M4AST online tool.\footnote{\url{http://spectre.imcce.fr/m4ast/index.php/index/home}} The tool fits a curve to the data and compares it to the taxons defined by \citet{2009Icar..202..160D} using a $\chi^2$ fitting procedure. The best three results are provided in order of decreasing goodness of fit. For the case of 2023~DZ$_{2}$, the best three fits are T, D, and Xe (see hatched region in Fig.~\ref{fig:DZ2spectrum}).

The T-types are defined by a linear spectrum with moderate to high slope at wavelengths bellow 0.75~$\mu$m  by both \citet{2002Icar..158..146B} and \citet{2009Icar..202..160D} taxonomies. Their spectral curves are within the range of D-types and X-complex asteroids (it is comparable with the reddest X-types and with the least red D-types). For this reason, we did not include this spectral type among the classes used for spectro-photometric classification.  The matching between the spectral data and the spectro-photometry covers the $r$, $i$, and $z_s$ bands (see Fig.~\ref{fig:tscvsgtc}). 

There is a slight discrepancy (at the level of  2-$\sigma_g$, where $\sigma_g$ is the standard deviation of the observations performed in the $g$ filter). There are several factors which can contribute to it, 1) the $g$ band covers the (0.4, 0.55)~$\mu$m spectral interval, while the spectrum stops at 0.48~$\mu$m; 2) the common wavelength range between 0.48 and 0.55~$\mu$m falls on the outermost boundary of the spectroscopic setup coverage, making it more susceptible to potential unaccounted errors in the slope; 3) errors that have not been taken into account could arise either from the photometric calibration or from the spectral data, such as issues similar to those reported by \cite{2020ApJS..247...73M} (although they refer to the near-infrared). With the existing data we cannot distinguish between these possibilities.

\section{Discussions and Conclusions\label{sec:conclusions}}

In the previous sections, we have presented a comprehensive analysis of ground-based observations of a close-approaching NEA, namely 2023~DZ$_{2}$. All these analyses would provide critical information for making decisions on mitigating strategies regarding a hypothetical impact (either on impact avoidance or damage limitation) should this NEA had any real impact risk. 
%In the previous sections, we have presented the critical steps needed to mitigate the threat linked to the cosmic hazard of SSO impacts. As an example, we used the case of 2023~DZ$_{2}$. First we show that continuous observations and real-time processing methods are required in order to detect potential impactors.

Our discovery of 2023~DZ$_{2}$ (initially classified as VI) with WFC mounted on INT, highlights the importance of continuous monitoring of the sky and of the need to process the data in near-real time. %Considering the wide sky coverage of the instrument and a cadence of 30--60 images (or more) per hour, the data reduction and the detection of new SSOs in the acquired CCD frames is not possible applying the classical methods. 
We showed that our ParaSOL infrastructure and the use of a synthetic tracking algorithm allowed us to process the large amount of data obtained and to report our findings in near-real time so the community could start contributing additional observation of the object of interest as early as possible. 

Once a potential impactor has been identified, it becomes critical to follow it up in order to improve its orbital determination and to obtain its physical properties, such as size, shape, composition, and structure. The estimation of size requires the knowledge of the absolute magnitude and the albedo \citep{1997Icar..126..450H}. The preliminary value of $H$ can be inferred from the apparent magnitude reported together with the astrometric measurements, after the orbit is computed. Nevertheless, this is a rough guess. The main source of the absolute magnitude errors is the irregular shape of the asteroids which generates light curves with amplitudes up to $\approx$2~mag, according to the current data existing in LCDB \citep{2009Icar..202..134W}. 

Thus, the photometric observations are the next step in characterizing the new NEA. In the case of 2023~DZ$_{2}$, we triggered observations with the TTT telescopes which revealed a rotation period of $P_{\rm rot}=6.2743\pm0.0005$~min and an amplitude of 0.57$\pm$ 0.14~mag. This result uncovered a fast-rotating object. The minimum aspect ratio for its shape was estimated from the light curve amplitude, $a/b = 1.30\pm0.08$ (where $a$ and $b$ are the axis perpendicular to the rotation axis). Again, this was an approximate determination because the three peaks in the light curve outlined a body with a complex shape. Because this is a fast rotator, this finding added a limit on the cohesive strength of the material it is made of, and allowed  to constrain the possible solutions for size and composition.

The value of the albedo is strongly dependent on the composition of the body \citep{2011AJ....142...85T,2011ApJ...743..156M,2018A&A...617A..12P}. It may vary in the interval of $\approx$(0.02, 0.50). In the case of 2023~DZ$_2$ and for the initial estimate of $H\approx$24~mag and this wide range of albedos, we can estimate a size between 30 to 150~m. %An improved determination could be obtained by performing a taxonomical classification (most taxonomic classes are characterized by a narrow interval of albedos).
%For the faintest objects, the most readily available technique for taxonomic classification is spectro-photometry, even if the obtained result is not as precise as the one obtained from spectroscopy. 
The albedo estimation can be made by assigning a taxonomic type. The most readily available technique for taxonomic classification is spectro-photometry, even if the obtained result is not as precise as the one obtained from spectroscopy. The observational results obtained with TCS allowed us to classify 2023 DZ$_2$ as an X-complex asteroid, the colour values $(g-r)=0.555\pm0.055$~mag, $(r-i)=0.1542\pm0.055$~mag, and $(i-z_s)=0.0638\pm0.059$~mag, place this object well inside the X-complex locus in the colour-colour diagram. 

%The reported errors are conservative and their value is limited by the photometric calibration (the errors of the zero points), the actual errors from the photometric reduction pipeline are of the order of milimagnitudes.

Unfortunately, the tentative classification as an X-complex type NEA does not constrain the albedo (this approach would have been successful for all the other classes). The X-complex includes both low and high albedo asteroids, with moderate spectral slope and featureless spectra, including various compositions similar to the carbonaceous, metallic and enstatite chondrites \citep{2011Icar..214..131F}. In the classification scheme of \cite{1984PhDT.........3T}, the  X-complex was divided in P -- primitives ($p_V\approx0.05$), M --  metallics ($p_V\approx0.015$), and E -- enstatites ($p_V\approx0.42$) classes.

With this information, there are several hints regarding the nature of 2023~DZ$_{2}$. Because it is a fast rotator, we show that it is highly unlikely to have a carbonaceous-like composition that corresponds to a dark albedo \citep{2023MNRAS.tmp..693L}. This consideration reduced the possible interval for its size to the range of 33  to  55~m (assuming an albedo of 0.42, and 0.15 respectively). Accurate spectra can confirm this classification and eventually reveal subtle features that may allow to discern its actual composition.

\begin{figure}[h!]
    \begin{center}
    \includegraphics[width=1\columnwidth]{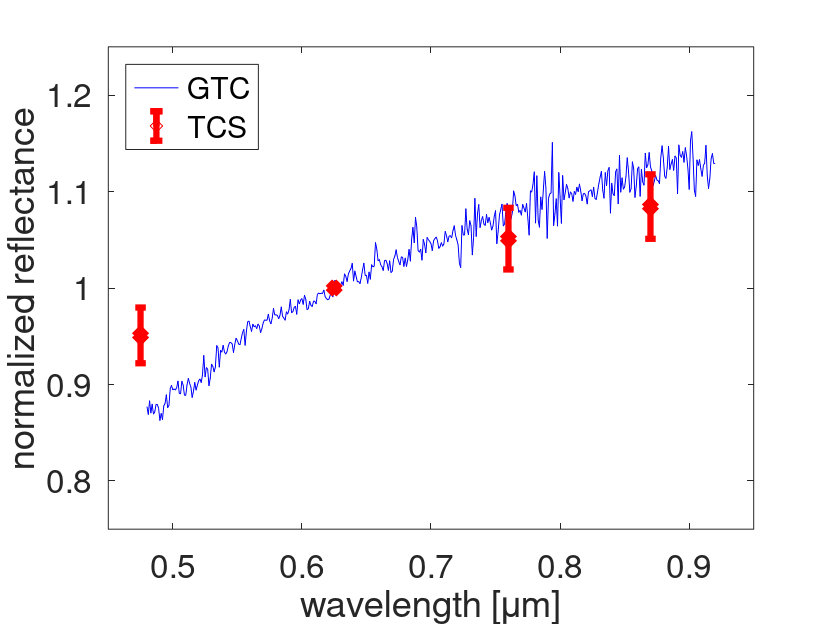}
    \caption{Comparison between the spectra of 2023~DZ$_{2}$ obtained with GTC and the spectro-photometric data obtained with TCS. In order to convert the TCS data, we used the following colours of the Sun, $(g-r)^{Sun}$ = 0.50~mag, $(r-i)^{Sun}$ = 0.10~mag, $(i-z_s)^{Sun}$ = 0.03~mag. These values (Popescu et al., in preparation) were derived for the filters available for the MuSCAT2 instrument and are consistent with those provided by \citet{2006MNRAS.367..449H}.}
    \label{fig:tscvsgtc}
    \end{center}
\end{figure}

However and in order to obtain the visible spectrum one must take into account that a 2.5~m class telescope is required for asteroids as faint as $V$ = 18.5~mag \citep[e.g.][]{2019A&A...627A.124P}, a 4~m class telescope is needed for objects as faint as $V$ = 20.5~mag \citep[e.g.][]{2018P&SS..157...82P}, and a 10~m class one, such as GTC --- the largest optical single mirror telescope currently available --- is required for bodies of $V\approx$ 22~mag \citep[e.g.][]{2023A&A...670L..10D}. Within our effort to characterize 2023~DZ$_{2}$, we used the 10.4~m GTC telescope during the nights of March 17 and March 20 to obtain spectra over the 0.46--0.92~${\mu}$m spectral range. The observations were performed when the object had respective apparent $V$ magnitudes of 19 and 18. The two spectral curves are identical and show a featureless spectrum with a moderate slope. The agreement between them and the spectro-photometric observations is shown in Fig.~\ref{fig:tscvsgtc}. 

%Between March 21 and March 27, 2023~DZ$_{2}$ was brighter than 18 magnitude (with the brightest apparent magnitude of $\approx$10). Thus, observers around the world obtained new data. The International Asteroid Warning Network (IAWN) \footnote{\url{https://iawn.net/}} organized a world wide campaign with the aim of involving as many observing facilities as possible , in a coordinated manner, for obtaining the most accurate physical information about this object.  This international organization was established in 2013 as a result of the UN-endorsed recommendations for an international response to a potential NEO impact threat. Their objective is to develop a "strategy using well-defined communication plans and protocols to assist Governments in the analysis of asteroid impact consequences and in the planning of mitigation responses."  While their previous IAWN campaigns were focused on the astrometric observations\citep{2022PSJ.....3..156F}, the close approach of 2023~DZ$_{2}$ offered a great opportunity for a world-wide collaboration to study a potential NEA impactor.

In this paper, we have described the first critical steps required to mitigate the threat of a potential impactor. We first detected 2023~DZ$_{2}$ using a near-real time processing algorithm, within the context of a limited survey in a region of the sky not covered by the large surveys. As a second step, less than seven days after the announcement of its discovery by the Minor Planet Center (on 16-March-2023), we were able to constrain its main physical properties using photometric observations (acquired a few days later, during March 20, 21, and 22), spectro-photometric (obtained on 22-March-2023) and spectroscopic data (observations performed on March 17 and 20). Regarding the origin of this NEA, its current orbit determination is not robust enough to find out which asteroid population is its most likely source; however, it is good enough to confirm that we are in no short-term danger of having a collision with 2023~DZ$_{2}$ thanks to the dynamical effects of a near secular (apsidal) resonance with Jupiter. This circumstance illustrates the dual nature of secular resonances, while $\nu_6$ might place asteroids in a collision course with Earth and the nodal resonance with Earth favours collisions, $\nu_5$ might help protecting Earth from some impacts. In the case of the object studied here, the protective secular near resonance effectively removes the risk connected with the nodal near resonance.

Although simulated scenarios for a potential impact were carried out during Planetary Defense Conferences in 2021 and 2023 \citep{2021plde.confE..57B}, and they were much more elaborated, here we confirm on a real case that key observations can be obtained within a few days after the announcement of a potential impact threat. This time the data collected eventually helped to confirm that the collision probability was insignificant, next time this may well not be the case: additional data may lead to reassuring a future impact instead. As the adage goes, "Time is of the essence when mitigating a cosmic hazard".

\begin{acknowledgements}
The work of MP, OV, MS, DB, LC and  MP,  was supported by a grant of the Romanian National Authority for  Scientific  Research -- UEFISCDI, project number PN-III-P2-2.1-PED-2021-3625. RdlFM and CdlFM thank S.~J. Aarseth for providing one of the codes used in this research and A.~I. G\'omez de Castro for providing access to computing facilities. This work was partially supported by the Spanish `Agencia Estatal de Investigaci\'on (Ministerio de Ciencia e Innovaci\'on)' under grant PID2020-116726RB-I00 /AEI/10.13039/501100011033. Based on observations made with the Isaac Newton Telescope (INT), in the Spanish Observatorio del Roque de los Muchachos of the Instituto de Astrof\'{\i}sica de Canarias (program ID INT99-MULTIPLE-2/23A). Based on observations made with the Gran Telescopio Canarias (GTC), installed at the Spanish Observatorio del Roque de los Muchachos of the Instituto de Astrof\'{\i}sica de Canarias, on the island of La Palma. This work is partly based on data obtained with the instrument OSIRIS, built by a Consortium led by the Instituto de Astrof\'{\i}sica de Canarias in collaboration with the Instituto de Astronom\'{\i}a of the Universidad Nacional Aut\'onoma de Mexico. OSIRIS was funded by GRANTECAN and the National Plan of Astronomy and Astrophysics of the Spanish Government.This paper includes observations made with the Two meter Twin Telescope (TTT) at the IAC’s Teide Observatory that Light Bridges, SL, operates on the Island of Tenerife, Canary Islands (Spain). The Observing Time Rights (DTO) used for this research at the TTT have been provided by the Instituto de Astrofísica de Canarias. The spectral and the spectro-photometric data were obtained in the framework of the European Union's Horizon 2020 research and innovation program under grant agreement No 870403 (NEOROCKS). JL, JdeL, M.R-A and MP acknowledge support from the ACIISI, Consejer\'{\i}a de Econom\'{\i}a, Conocimiento y Empleo del Gobierno de Canarias and the European Regional Development Fund (ERDF) under grant with reference ProID2021010134. In preparation of this paper, we made use of the NASA Astrophysics Data System, the ASTRO-PH e-print server, and the MPC data server.  We thank the reviewers for their comments which helped us to improve the paper. We also thank to Dr. Hissa Medeiros for the discussions on this topic and for the suggested references, to Gabriel Nicolae Simon for providing the python functions used for the spectro-photometric classification. 
\end{acknowledgements}

\bibliographystyle{aa}      
\bibliography{2023DZ2_Paper}   % name your BibTeX data base

\begin{appendix}
\section{Pro - Am collaborations for photometric characterization}

Events that require rapid response observations or long-term monitoring campaigns of diverse celestial sources are the triggers for the collaboration between professional and amateur astronomers who have access to highly reliable observing facilities\footnote{\url{https://www.nature.com/articles/s41550-023-01954-6}}. During its close approach, 2023~DZ$_{2}$ was bright enough for observations with small aperture telescopes. Thus, the light curves obtained initially could be followed-up by many amateur astronomers.

\begin{figure*}[h!]
    \begin{center}
    \includegraphics[width=11cm]{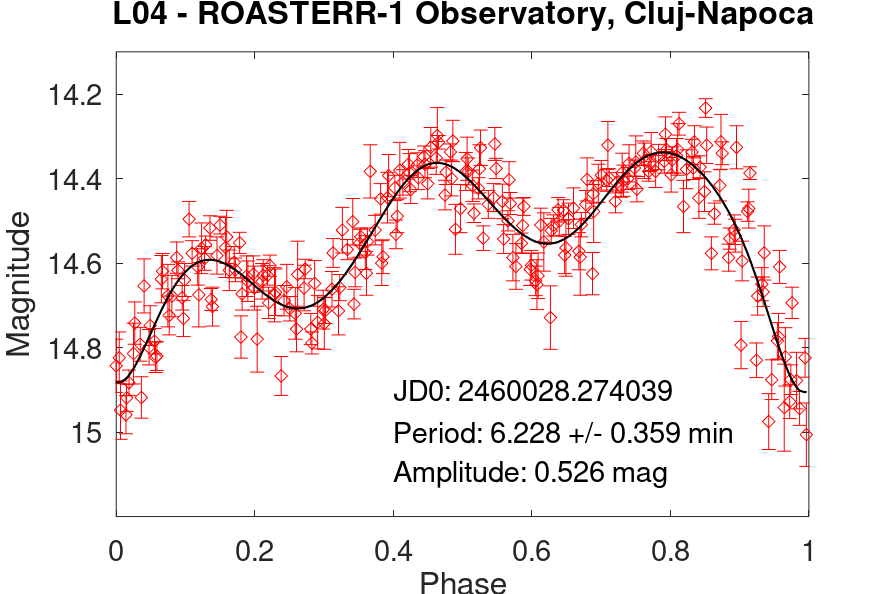}
    \includegraphics[width=11cm]{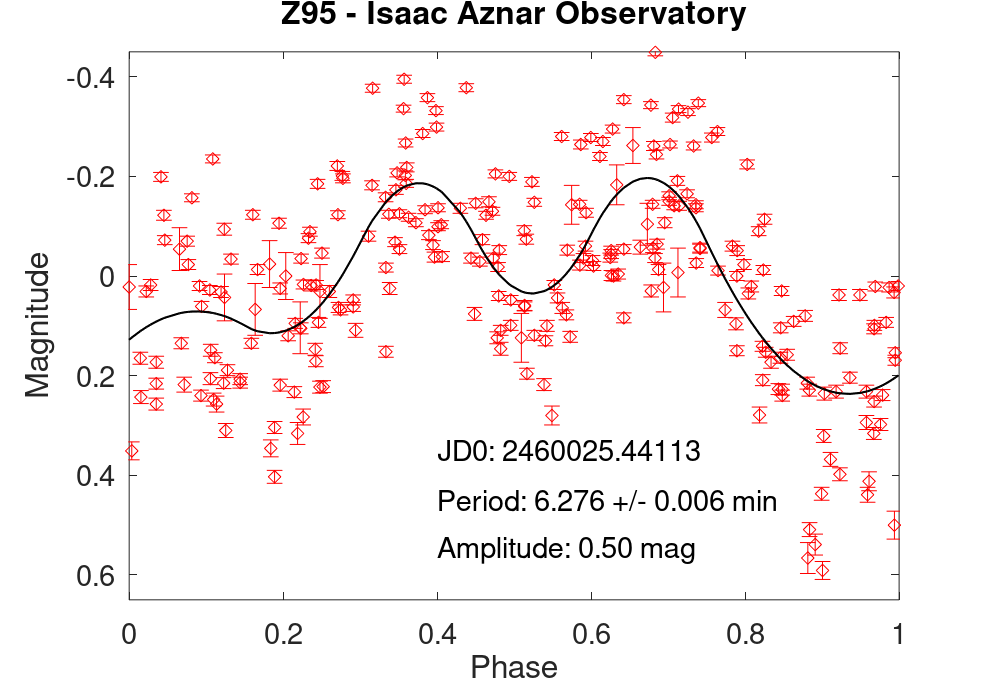}
    \includegraphics[width=11cm]{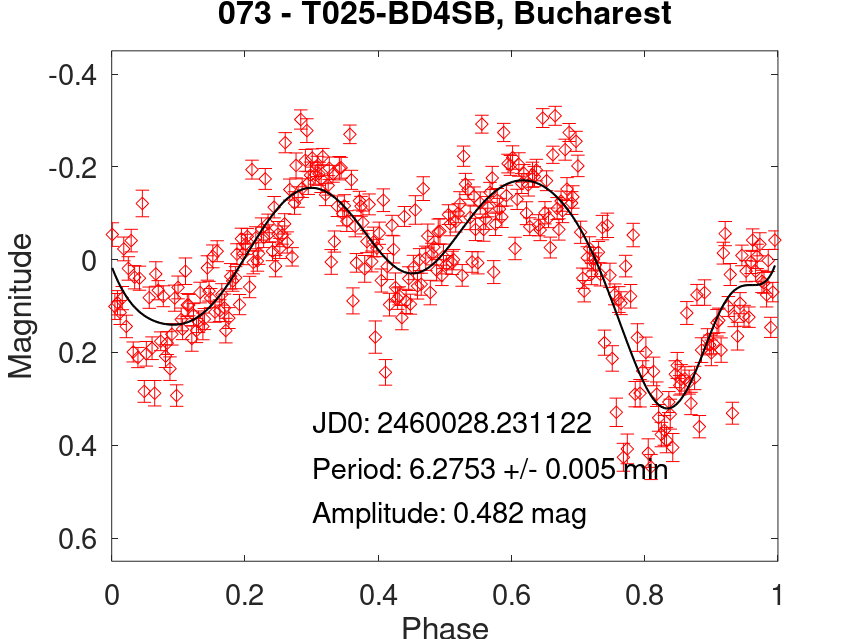}
    \caption{Phased light curve of 2023~DZ$_{2}$ computed using photometric measurements obtained within the professional--amateurs collaboration discussed in the text.  The title of each panel highlights the observatory (including the MPC code). The reported period and amplitude were computed by the observers using {\tt Tycho Tracker} (for \emph{L04} and \emph{073}), and {\tt MPOC Canopus} (for \emph{Z95}). The plots are made with GNU Octave software \citep{octave}, and a splinefit function was used to fit the data. In the case of 073, a binning of nine points was made after folding the photometric data.}
    \label{fig:proam}
    \end{center}
\end{figure*}

We received various contributions from different amateurs from all over the world, including Amadeo Aznar, Lucian Hudin, and Cristian Suciu (Astroclubul Bucure\c{s}ti, Romania). To highlight the importance of Pro-Am collaborations, we present here some of the most reliable light curves we received (Fig.~\ref{fig:proam}) and obtained by the Isaac Aznar Observatory (IAO) Alcublas (MPC code Z95), Valencia province, Spain;  ROASTERR-1 Observatory (MPC code L04), Cluj-Napoca, Romania; and T025-BD4SB (MPC code 073), Bucharest, Romania \citep{2022EPSC...16.1222N}. We choose to present here the typical plots which can be obtained using commercially available tools such as {\tt MPOC Canopus} and {\tt Tycho Tracker}. With help from these tools, experienced amateur astronomers can derive the spin properties of asteroids.    

Photometric observations of 2023~DZ$_{2}$  were performed from the ROASTERR-1 observatory with the 0.3~m f/5 corrected Newtonian, a CCD camera with a KAF-8300 CCD chip cooled at -15$^o$C, and a clear filter. The calibration data were acquired immediately after the observations and consisted of 32 bias, 32 dark, and 32 flat frames. Two sets of observations were used, made of 104 images and 138 images, respectively, with an exposure time of 10~s for each individual image. The {\tt Tycho Tracker} v10 was used for photometry. The obtained period was 0.1038~h (6.228~min) with an amplitude of 0.526~mag as shown in Fig.~\ref{fig:proam}. 

The Isaac Aznar Observatory (IAO) is privately owned by the Spanish amateur astronomer Amadeo Aznar Mac\'{\i}as. It is located in Alcublas, Valencia province, at 900~m above the sea level, in one of the darkest night-sky of the Iberian Peninsula (limiting magnitude 21.8 mag~arcsec$^{-2}$). The optical system consists of a remotely controlled 0.36~m Schmidt-Cassegrain F/10 Meade LX200 telescope. The CCD camera is a SBIG STL 11000, with 1.68~arcsec~pixel$^{-1}$ resolution (pixel binned $\times$3) and a square FoV of 37$\times$24~arcmin$^2$. The light curve of 2023~DZ$_{2}$ was obtained using 274 photometric exposures from two nights (see Fig.~\ref{fig:proam}), on March 21 and 25. An exposure time of 30~s was used, and images were acquired during $\approx$2.2~h. The data reduction was performed using the {\tt MPO Canopus} software.\footnote{\url{https://minplanobs.org/BdwPub/php/displayhome.php}} The light curve shows a rotation period of 0.1046~h (6.276~min) with an error of 0.0001~h and an amplitude of 0.50~mag with an error of 0.08~mag.

The T025–BD4SB robotic telescope was built as a collaboration between amateur astronomers from the Astroclubul Bucure\c{s}ti and the researchers from the Astronomical Institute of the Romanian Academy. The main components of this instrument are a 10~inch Newtonian telescope and a QHY 294M CMOS camera. T025–BD4SB is mounted on the roof of the old Bucharest Astronomical Institute building. This facility has the Minor Planet Center observatory code 073. The detection limit is around $V$=20~mag, and the median seeing is 2.8 arcsec. The observations of  2023~DZ$_{2}$ were performed on the night of 24-March-2023 (one night prior to the close approach between 2023~DZ$_{2}$ and Earth). Continuous exposures of 5~s each were acquired during $\approx$5~h. This short exposure time was considered because of its high apparent motion (30--40 arcsec per min during the observations). By using this exposure time, the trail left on the image by the NEA matches the typical seeing from Bucharest. The data reduction followed the same steps as those described above. The image pre-processing was carried out using the Pyraf package.\footnote{\url{https://iraf-community.github.io/pyraf.html}} Because of the light-pollution of Bucharest, a sky-background removal algorithm was applied using the GNU Astronomy Utilities (Gnuastro) package \citep{gnuastro}. The results obtained with the T025-BD4SB, were a rotation period $P_{\rm rot}=6.2753\pm0.005$~min and an amplitude of 0.482~mag.

      \section{Input data\label{Adata}}
         Here, we include the barycentric Cartesian state vector of NEA 2023~DZ$_{2}$. This vector and its uncertainties have been used to 
         perform the calculations discussed above. For example, a new value of the $X$ component of the state vector is computed as $X_{\rm c} = X + \sigma_X \ r$, 
         where $r$ is an univariate Gaussian random number, and $X$ and $\sigma_X$ are the mean value and its 1$\sigma$ uncertainty in  
         Table~\ref{vector2023DZ2}.
%
%---------------------------------------------------------------------------------------------------------------------------------- TABLE II
%------------------------------------------------------------------------------------------------- Geometric Cartesian state vector 2023 DZ2
%
     \begin{table}
      \centering
      \fontsize{8}{12pt}\selectfont
      \tabcolsep 0.15truecm
      \caption{\label{vector2023DZ2}Barycentric Cartesian state vector of 2023~DZ$_{2}$: components and associated 1$\sigma$ uncertainties.
              }
      \begin{tabular}{ccc}
       \hline
        Component                         &   &    value$\pm$1$\sigma$ uncertainty                                 \\
       \hline
        $X$ (au)                          & = & $-$9.718051175253550$\times10^{-1}$$\pm$1.99342196$\times10^{-8}$  \\
        $Y$ (au)                          & = &    5.157636594173508$\times10^{-1}$$\pm$9.69360581$\times10^{-9}$     \\
        $Z$ (au)                          & = & $-$7.025348786300637$\times10^{-4}$$\pm$1.18266109$\times10^{-8}$  \\
        $V_X$ (au/d)                      & = &  $-$4.772825832549133$\times10^{-3}$$\pm$8.13837751$\times10^{-10}$  \\
        $V_Y$ (au/d)                      & = & $-$1.954511508654970$\times10^{-2}$$\pm$2.94881778$\times10^{-10}$  \\
        $V_Z$ (au/d)                      & = &      
           2.660147805349922$\times10^{-5}$$\pm$3.86774841$\times10^{-10}$  \\
       \hline  
      \end{tabular}
      \tablefoot{Data are referred to epoch JD 2460000.5, which corresponds to 0:00 on 25-February-2023, TDB (J2000.0 ecliptic and equinox). Source: JPL's {\tt Horizons}.
                }
     \end{table}
%
%-------------------------------------------------------------------------------------------------------------------------------------------
%

\end{appendix}

\end{document}